\documentclass[aps,prd,twocolumn,showpacs,nofootinbib,floatfix,superscriptaddress]{revtex4}

\usepackage{amsmath,bm}
\usepackage{amssymb}
\usepackage{mathrsfs}
\usepackage{color}
\usepackage{dcolumn}
\usepackage{graphicx}
\usepackage{multirow}
\usepackage{rotating}
\usepackage{afterpage}
\usepackage{enumerate}

\usepackage{xspace}
\usepackage{url}

\usepackage{fancyhdr}
\usepackage{float}
\usepackage{psfrag}

\newcommand{\herwig}    {\textsc{herwig}\xspace}
\newcommand{\pythia}    {\textsc{pythia}\xspace}
\newcommand{\alpgen}    {\textsc{alpgen}\xspace}

\newcommand{\geant}     {\textsc{geant3}\xspace}
\newcommand{\mcatnlo}   {\textsc{mc@nlo}\xspace}
\newcommand{\resbos}   {\textsc{resbos}\xspace}

\newcommand{\alppyt}    {{\sc alpgen+pythia}\xspace}
\newcommand{\alpher}    {{\sc alpgen+herwig}\xspace}
\newcommand{\mcherwig}  {{\sc mc@nlo+herwig}\xspace}
\newcommand{\comphep}   {{\sc comphep}\xspace}
\newcommand{\mcfm}   {{\sc mcfm}\xspace}
\newcommand{\tunfold}   {{\sc tunfold}\xspace}

\newcommand{\met}       {\mbox{$\not\!\!E_T$}\xspace}
\newcommand{\metNo}       {\mbox{$\not\!\!E_T$}}

\newcommand{\wplus}     {\ensuremath{W+}jets\xspace}
\newcommand{\zplus}     {\ensuremath{Z/\gamma^{*}+}jets\xspace}
\newcommand{\ptmiss}    {\ensuremath{{p\kern-0.5em\slash}_{T}}\xspace}
\newcommand{\muplus}    {\ensuremath{\mu +}jets\xspace}
\newcommand{\eplus}     {$e +$jets\xspace}

\newcommand{\lplustw}     {$\ell +2$\,jets\xspace}
\newcommand{\lplusth}     {$\ell +3$\,jets\xspace}

\newcommand{\lplusgefo}     {$\ell + \ge 4$\,jets\xspace}

\newcommand{\ppbar}{\ensuremath{p\bar{p}}\xspace}
\newcommand{\qqbar}{\ensuremath{q\bar{q}}\xspace}
\newcommand{\ttbar}{\ensuremath{t\bar{t}}\xspace}
\newcommand{\mTT}{\ensuremath{m(t\bar{t})}\xspace}
\newcommand{\ptTT}{\ensuremath{p_{T}^{t\bar{t}}}\xspace}
\newcommand{\ptt}{\ensuremath{p_{T}^{\mathrm{top}}}\xspace}

\newcommand{\aetat}{\ensuremath{|y^{\mathrm{top}}|}\xspace}
\newcommand{\wjets}{\ensuremath{W}+\rm jets\xspace}
\newcommand{\zjets}{\ensuremath{Z/\gamma^{*}}+\rm jets\xspace}
\newcommand{\mm}{\mathrm}

\newcommand{\ljets}{$\ell +$jets\xspace}
\newcommand{\dzero}     {D0\xspace}

\newcommand\T{\rule{0pt}{2.6ex}}       % Top strut

\newcolumntype{d}[1]{D{.}{.}{#1}}
\begin{document}

% remove all headers from document
\pagestyle{plain}

% remove the space for publication
% \vspace*{1.5cm}

\title{Measurement of differential {\boldmath \ttbar} production cross sections in 
{\boldmath \ppbar} collisions}

%---------------------------------------------------------------------

\hspace{5.2in} \mbox{FERMILAB-PUB-14-012-E}

%%%%%%%%%%%%%%%%%%%%%%%%%%%%%%%%%%%%%%%%%%%%%%%%%%%%%%%%%%

%
\affiliation{LAFEX, Centro Brasileiro de Pesquisas F\'{i}sicas, Rio de Janeiro, Brazil}
\affiliation{Universidade do Estado do Rio de Janeiro, Rio de Janeiro, Brazil}
\affiliation{Universidade Federal do ABC, Santo Andr\'e, Brazil}
\affiliation{University of Science and Technology of China, Hefei, People's Republic of China}
\affiliation{Universidad de los Andes, Bogot\'a, Colombia}
\affiliation{Charles University, Faculty of Mathematics and Physics, Center for Particle Physics, Prague, Czech Republic}
\affiliation{Czech Technical University in Prague, Prague, Czech Republic}
\affiliation{Institute of Physics, Academy of Sciences of the Czech Republic, Prague, Czech Republic}
\affiliation{Universidad San Francisco de Quito, Quito, Ecuador}
\affiliation{LPC, Universit\'e Blaise Pascal, CNRS/IN2P3, Clermont, France}
\affiliation{LPSC, Universit\'e Joseph Fourier Grenoble 1, CNRS/IN2P3, Institut National Polytechnique de Grenoble, Grenoble, France}
\affiliation{CPPM, Aix-Marseille Universit\'e, CNRS/IN2P3, Marseille, France}
\affiliation{LAL, Universit\'e Paris-Sud, CNRS/IN2P3, Orsay, France}
\affiliation{LPNHE, Universit\'es Paris VI and VII, CNRS/IN2P3, Paris, France}
\affiliation{CEA, Irfu, SPP, Saclay, France}
\affiliation{IPHC, Universit\'e de Strasbourg, CNRS/IN2P3, Strasbourg, France}
\affiliation{IPNL, Universit\'e Lyon 1, CNRS/IN2P3, Villeurbanne, France and Universit\'e de Lyon, Lyon, France}
\affiliation{III. Physikalisches Institut A, RWTH Aachen University, Aachen, Germany}
\affiliation{Physikalisches Institut, Universit\"at Freiburg, Freiburg, Germany}
\affiliation{II. Physikalisches Institut, Georg-August-Universit\"at G\"ottingen, G\"ottingen, Germany}
\affiliation{Institut f\"ur Physik, Universit\"at Mainz, Mainz, Germany}
\affiliation{Ludwig-Maximilians-Universit\"at M\"unchen, M\"unchen, Germany}
\affiliation{Panjab University, Chandigarh, India}
\affiliation{Delhi University, Delhi, India}
\affiliation{Tata Institute of Fundamental Research, Mumbai, India}
\affiliation{University College Dublin, Dublin, Ireland}
\affiliation{Korea Detector Laboratory, Korea University, Seoul, Korea}
\affiliation{CINVESTAV, Mexico City, Mexico}
\affiliation{Nikhef, Science Park, Amsterdam, The Netherlands}
\affiliation{Radboud University Nijmegen, Nijmegen, The Netherlands}
\affiliation{Joint Institute for Nuclear Research, Dubna, Russia}
\affiliation{Institute for Theoretical and Experimental Physics, Moscow, Russia}
\affiliation{Moscow State University, Moscow, Russia}
\affiliation{Institute for High Energy Physics, Protvino, Russia}
\affiliation{Petersburg Nuclear Physics Institute, St. Petersburg, Russia}
\affiliation{Instituci\'{o} Catalana de Recerca i Estudis Avan\c{c}ats (ICREA) and Institut de F\'{i}sica d'Altes Energies (IFAE), Barcelona, Spain}
\affiliation{Uppsala University, Uppsala, Sweden}
\affiliation{Taras Shevchenko National University of Kyiv, Kiev, Ukraine}
\affiliation{Lancaster University, Lancaster LA1 4YB, United Kingdom}
\affiliation{Imperial College London, London SW7 2AZ, United Kingdom}
\affiliation{The University of Manchester, Manchester M13 9PL, United Kingdom}
\affiliation{University of Arizona, Tucson, Arizona 85721, USA}
\affiliation{University of California Riverside, Riverside, California 92521, USA}
\affiliation{Florida State University, Tallahassee, Florida 32306, USA}
\affiliation{Fermi National Accelerator Laboratory, Batavia, Illinois 60510, USA}
\affiliation{University of Illinois at Chicago, Chicago, Illinois 60607, USA}
\affiliation{Northern Illinois University, DeKalb, Illinois 60115, USA}
\affiliation{Northwestern University, Evanston, Illinois 60208, USA}
\affiliation{Indiana University, Bloomington, Indiana 47405, USA}
\affiliation{Purdue University Calumet, Hammond, Indiana 46323, USA}
\affiliation{University of Notre Dame, Notre Dame, Indiana 46556, USA}
\affiliation{Iowa State University, Ames, Iowa 50011, USA}
\affiliation{University of Kansas, Lawrence, Kansas 66045, USA}
\affiliation{Louisiana Tech University, Ruston, Louisiana 71272, USA}
\affiliation{Northeastern University, Boston, Massachusetts 02115, USA}
\affiliation{University of Michigan, Ann Arbor, Michigan 48109, USA}
\affiliation{Michigan State University, East Lansing, Michigan 48824, USA}
\affiliation{University of Mississippi, University, Mississippi 38677, USA}
\affiliation{University of Nebraska, Lincoln, Nebraska 68588, USA}
\affiliation{Rutgers University, Piscataway, New Jersey 08855, USA}
\affiliation{Princeton University, Princeton, New Jersey 08544, USA}
\affiliation{State University of New York, Buffalo, New York 14260, USA}
\affiliation{University of Rochester, Rochester, New York 14627, USA}
\affiliation{State University of New York, Stony Brook, New York 11794, USA}
\affiliation{Brookhaven National Laboratory, Upton, New York 11973, USA}
\affiliation{Langston University, Langston, Oklahoma 73050, USA}
\affiliation{University of Oklahoma, Norman, Oklahoma 73019, USA}
\affiliation{Oklahoma State University, Stillwater, Oklahoma 74078, USA}
\affiliation{Brown University, Providence, Rhode Island 02912, USA}
\affiliation{University of Texas, Arlington, Texas 76019, USA}
\affiliation{Southern Methodist University, Dallas, Texas 75275, USA}
\affiliation{Rice University, Houston, Texas 77005, USA}
\affiliation{University of Virginia, Charlottesville, Virginia 22904, USA}
\affiliation{University of Washington, Seattle, Washington 98195, USA}
\author{V.M.~Abazov} \affiliation{Joint Institute for Nuclear Research, Dubna, Russia}
\author{B.~Abbott} \affiliation{University of Oklahoma, Norman, Oklahoma 73019, USA}
\author{B.S.~Acharya} \affiliation{Tata Institute of Fundamental Research, Mumbai, India}
\author{M.~Adams} \affiliation{University of Illinois at Chicago, Chicago, Illinois 60607, USA}
\author{T.~Adams} \affiliation{Florida State University, Tallahassee, Florida 32306, USA}
\author{J.P.~Agnew} \affiliation{The University of Manchester, Manchester M13 9PL, United Kingdom}
\author{G.D.~Alexeev} \affiliation{Joint Institute for Nuclear Research, Dubna, Russia}
\author{G.~Alkhazov} \affiliation{Petersburg Nuclear Physics Institute, St. Petersburg, Russia}
\author{A.~Alton$^{a}$} \affiliation{University of Michigan, Ann Arbor, Michigan 48109, USA}
\author{A.~Askew} \affiliation{Florida State University, Tallahassee, Florida 32306, USA}
% ...
%
% visitor_addresses.tex                       28 December 2013
%  available symbols are:
%  $\ast, \dag, \ddag, \S, \P, $\|$, $\ast\ast$, \dag\dag, \ddag\ddag ,\#
%
\collaboration{The D0 Collaboration\footnote{with visitors from
%{alton}
$^{a}$Augustana College, Sioux Falls, SD, USA,
%{burdin}
$^{b}$The University of Liverpool, Liverpool, UK,
%{grohsjean}
$^{c}$DESY, Hamburg, Germany,
%{de la cruz-burelo}
$^{d}$Universidad Michoacana de San Nicolas de Hidalgo, Morelia, Mexico
%{partridge}
$^{e}$SLAC, Menlo Park, CA, USA,
%{hesketh}
$^{f}$University College London, London, UK,
%{luna-garcia}
$^{g}$Centro de Investigacion en Computacion - IPN, Mexico City, Mexico,
%{santos}
$^{h}$Universidade Estadual Paulista, S\~ao Paulo, Brazil,
%{meyer}
$^{i}$Karlsruher Institut f\"ur Technologie (KIT) - Steinbuch Centre for Computing (SCC),
D-76128 Karlsrue, Germany,
%{patwa}
$^{j}$Office of Science, U.S. Department of Energy, Washington, D.C. 20585, USA,
%{cooke}
$^{k}$American Association for the Advancement of Science, Washington, D.C. 20005, USA
and
%{borysova}
$^{l}$Kiev Institute for Nuclear Research, Kiev, Ukraine
$^{m}$ETH Z\u rich, Z\u rich, Switzerland 
%{montgomery}
%$^{?}$Thomas Jefferson National Accelerator Facility, Newport News, VA 23606, USA,
%{falkowski}
%$^{?}$Laboratoire de Physique Theorique, Orsay, FR,
%{hooper,kozminski}
%$^{?}$}Visitor from Lewis University, Romeoville, IL, USA.
%{weber}
%$^{?}$Universit{\"a}t Bern, Bern, Switzerland.
%{deceased}
%{zanabria}
%$^{?}$City Colleges of Chicago, Chicago, IL, USA}
%$^{\ddag}$Deceased.
}} \noaffiliation
\vskip 0.25cm

\author{}
\date{01/22/2014}

%---------------------------------------------------------------------

\begin{abstract}

The production of top quark-antiquark pair events in \ppbar collisions at $\sqrt{s}=1.96$ TeV is studied as a function of the transverse momentum and absolute value of the rapidity of the top quarks as well as of the invariant mass of the \ttbar pair. We select events containing an isolated lepton, a large imbalance in transverse momentum, and four or more jets with at least one jet identified as originating from a $b$ quark. The data sample corresponds to 9.7 fb${}^{-1}$ of integrated luminosity recorded with the \dzero detector during Run II of the Fermilab Tevatron Collider. Observed differential cross sections are consistent with standard model predictions.

\end{abstract}


\maketitle 

%---------------------------------------------------------------------

% \modulolinenumbers[1]
% \linenumbers

%%%%%%%%%%%%%%%%%%%%%%%%%%%%%%%%%%%%%%%%%%%%%%%%%%%%%%%%%%%%%%%%%%%%%%%%
%%%%%%%%%%%%%%%%%%%%%%%%%%%%%%%%%%%%%%%%%%%%%%%%%%%%%%%%%%%%%%%%%%%%%%%%
%%%%%%%%%%%%%%%%%%%%%%%%%%%%%%%%%%%%%%%%%%%%%%%%%%%%%%%%%%%%%%%%%%%%%%%%

\section{Introduction\label{toc:Intro}}
The top quark, discovered by the CDF and \dzero experiments in 1995 \cite{top_disc1,top_disc2}, is the heaviest of all elementary particles in the standard model (SM), with a mass of $173.2 \pm 0.9~\mm{GeV}$ \cite{TevatronMassCombo}. The production of top quark-antiquark pairs (\ttbar) at the Fermilab Tevatron Collider is dominated by the quark-antiquark (\qqbar) annihilation process. The measurement of \ttbar differential production cross sections provides a direct test of quantum chromodynamics (QCD), the theory of the strong interactions. %Measurements of differential cross sections deepen our understanding of QCD, and provide important information that can improve the simulation of QCD processes.
Moreover, a precise modeling of QCD processes is vital in many searches for contributions from new phenomena, where differential top quark cross sections can be used to set constraints on new sources of physics. A detailed understanding of top quark production is also needed for measurements or searches where new particles decay to a \ttbar pair, where other particles are produced in association with a \ttbar pair, or where \ttbar production is among the dominant backgrounds. An example of the importance of accurate modeling of QCD is given by the deviation observed in the charge asymmetry measurement in $p\bar{p} \rightarrow t\bar{t}$ production from SM predictions \cite{d0_afb,cdf_afb,d0_afb_lep,cdf_afb_lep}. Such a difference could be due to the exchange of a new heavy mediator, e.g., an axigluon \cite{axi1,axi2} that could also enhance the \ttbar cross section. Differential cross sections, most notably the one as a function of the invariant mass of the \ttbar pair $d\sigma/dm(t\bar{t})$, provide stringent constraints on axigluon models \cite{cdfmtt}. Differential \ttbar production cross sections have been previously measured at both the Tevatron \cite{dzero_ptt,cdfmtt} and the LHC \cite{cms_diff,atlas_diff}. The earlier measurements of differential \ttbar production at the Tevatron as a function of the transverse momentum of the $t$ and $\bar{t}$ quark (\ptt) \cite{dzero_ptt}, and as a function of \mTT \cite{cdfmtt}, showed good agreement with perturbative QCD (pQCD) calculations at next-to-leading (NLO), as well as next-to-next-to-leading order (NNLO) \cite{ptt_nnlo}. Compared to the previous \dzero result \cite{dzero_ptt}, the current measurement employs a factor of 10 more data allowing for higher precision tests of pQCD.\\

Single differential cross sections are measured as a function of \mTT, the absolute value of the rapidity\footnote{The rapidity $y$ is defined as $y = 1/2 \cdot \ln[ (E+p_z)/(E-p_z)]$, where $E$ is the energy of a particle and $p_z$ is the $z$ component of its momentum vector $\vec{p}$. The direction of the $z$ axis is defined along the proton beam direction.} \aetat, and \ptt, using events with a topology consistent with \ttbar decays. The index ``top" in \aetat and \ptt refers to either $t$ or $\bar{t}$ quarks. The observed $t$ and $\bar{t}$ differential distributions are consistent with each other, hence they are combined. Events are selected in the lepton+jets decay channel, where the lepton ($\ell$) refers to either an electron or a muon. This channel corresponds to $t\bar{t} \rightarrow W^{+}b W^{-}\bar{b}$ decays, where one of the two $W$ bosons decays leptonically ($W \rightarrow \ell \nu$), and the other hadronically ($W \rightarrow q\bar{q}'$). This decay channel includes also small contributions from electrons and muons stemming from the decay of $\tau$ leptons ($t \rightarrow Wb \rightarrow \tau \nu_{\tau} b \rightarrow \ell \nu_{\ell} \nu_{\tau} b$). The events are required to contain, in addition to the lepton, at least four jets and an imbalance in transverse momentum \met, as discussed in Sec.\ \ref{toc:data_samples}.

%***********************************************************
%
%
%
%
%
\section{Monte Carlo Simulations and QCD predictions}
\label{toc:generators}
Monte Carlo (MC) simulations are used to model the reconstruction of the observables, to estimate systematic uncertainties associated with the measurements, and to simulate physics processes. Different MC event generators are used to implement hard scattering processes based on leading-order (LO) and NLO QCD calculations, and are complemented with parton shower evolution. To simulate detector effects, generated events (including hadronization) are passed through a detailed simulation of the D0 detector response based on \geant \cite{geant}. To account for effects from additional overlapping \ppbar interactions, %``zero bias" events are selected randomly in collider data and overlaid on the fully simulated MC events.
 events without any trigger requirements are selected randomly in collider data and overlaid on the fully simulated MC events.

The \ttbar samples are generated with \mcatnlo version 3.4 \cite{mcatnlo}, which includes the production of off-shell top quarks by taking into account their finite width or with \alpgen version 2.11 \cite{alpgen}, which produces only on-shell top quarks. Single top quark production $(q\bar{q}' \rightarrow t\bar{b}, q'g \rightarrow tq\bar{b})$ is modeled using \comphep \cite{comphep}. For events generated with \mcatnlo, the parton showering is performed with \herwig version 6.510 \cite{herwig}, whereas for \alpgen and \comphep parton showering is implemented by \pythia version 6.409 \cite{pythia}. In the following the term ``scale" and the symbol $\mu$ refer to the renormalization and factorization scales, which are assumed to be equal and evaluated for the specific processes. The parton density functions (PDF), and other choices made in generating simulated events are summarized in Table \ref{tab:FacRenScales}. For all the MC simulations involving the generation of top quarks a top quark mass of \mbox{$m_t = 172.5$ GeV} is used. The difference from the current Tevatron top quark mass measurement of \mbox{173.2 GeV} \cite{TevatronMassCombo} has negligible impact on the analysis and is treated as a systematic uncertainty (see Sec.\ \ref{toc:xsec_sys}).

\begin{table}[htdp]
\begin{center}
    \caption{\label{tab:FacRenScales} Details of the signal and background modeling employed in this measurement. All final-state particles are used to compute the chosen scale, except the decay products of the $W$ boson, and are consequently used to calculate the mass $m$ and $p_T$. The term $m_V$ refers to the mass of the $W$ or $Z$ boson. The CTEQ6L1 \cite{cteq6l} and CTEQ6M \cite{cteq6m} PDFs are used.
}
\begin{ruledtabular}
\begin{tabular}{l c c c}
Process & Generator & Scale, $\mu$ & PDF \T \\ \hline
\ttbar &\alpgen & $\sqrt{\sum(m^2 + p_T^2)}$ & CTEQ6L1 \T \\
\ttbar &\mcatnlo & $\sqrt{\sum(m^2 + p_T^2)}$ & CTEQ6M \\
\wplus 	  &  \alpgen & $\sqrt{m_V^2 + \sum(m^2 + p_T^2)}$ & CTEQ6L1 \\
\zplus	  &  \alpgen & $\sqrt{m_V^2 + \sum(m^2 + p_T^2)}$ & CTEQ6L1 \\
Diboson	          &  \pythia & $\sqrt{m_V^2 + \sum(m^2 + p_T^2)}$ & CTEQ6L1 \\
Single top  & \comphep & $m_t$  & CTEQ6L1\\
~($s$ channel)  & & & \\
Single top  & \comphep & $m_t/2$ & CTEQ6M \\
~($t$ channel)  & & & \\
\end{tabular}
\end{ruledtabular}
\end{center}
\end{table}%

Several QCD predictions for differential \ttbar cross sections have been calculated at higher orders than those included in the MC generators. They use approximate NNLO calculations based on next-to-next-to-leading logarithm (NNLL) resummation for \mbox{$m_t = 173$ GeV} to calculate the \ptt and \aetat differential distributions \cite{ptt_nnlo,etat_nnlo}, and \mbox{$m_t = 172.5$ GeV} to calculate the \mTT and \ptt differential distributions \cite{mtt_nnlo}. All use the MSTW2008NNLO PDF \cite{mstw2008nnlo}. The scale used to calculate the \ptt and \aetat differential distributions is $m_t$. Employing $m_t$ as the scale for calculating the \mTT distribution leads to large and negative NLO corrections that result in negative differential cross sections at approximate NNLO, especially at large \mTT. In Ref.\ \cite{mtt_nnlo}, the \mTT distribution is calculated using the scale \mTT instead, which avoids this issue, but leads to a $7.7 \%$ lower inclusive cross section.\\
When comparing to \dzero data, we normalize the total cross section of the calculations in Ref.\ \cite{mtt_nnlo} for the \ptt and \mTT distributions to match the inclusive fully resummed NNLL at NNLO QCD calculation (using $m_t = 172.5$ GeV and the MSTW2008NNLO PDF), which finds \mbox{$\sigma_{\mathrm{tot}}^{\mathrm{res}} = 7.35 ^{+0.23}_{-0.27}\thinspace(\mathrm{scale} + \mathrm{pdf})$ pb} \cite{nnloInclXsec}. The total cross section of the approximate NNLO calculation as in Ref.\ \cite{ptt_nnlo, etat_nnlo} is calculated from the \ptt distribution and yields \mbox{$7.08 ^{+0.20}_{-0.24}\thinspace(\mathrm{scale}) ^{+0.36}_{-0.27} (\mathrm{PDF})$ pb}. The inclusive cross section calculated by integrating the \aetat or \ptt distribution deviates by 1.1\%. %A very similar value is calculated when integrating the \aetat distribution. %This calculation is, in contrast to the calculations in Ref.\ \cite{mtt_nnlo}, not re-normalized to match the fully resummed NNLL at NNLO QCD.
For reasons of consistency, the \ptt and \aetat distributions from Refs.\ \cite{ptt_nnlo, etat_nnlo} are not rescaled from their original predictions.

\subsection{Backgrounds}
The main background to \ttbar in the \ljets final state is \wplus production. It consists of events where one $W$ boson is produced via an electroweak interaction, together with additional partons from QCD processes. The \wplus final state can be split into four subsamples according to parton flavor: $Wb\bar{b}+\mm{jets}$, $Wc\bar{c}+\mm{jets}$, $Wc+\mm{jets}$ and $W+$light jets, where light refers to gluons, $u$, $d$ or $s$ quarks. The LO \alpgen cross sections are corrected for NLO effects as provided by \mcfm \cite{mcfm}: the $W+\mm{jets}$ cross section is multiplied by 1.30, and the $Wb\bar{b}+\mm{jets}$ and $Wc\bar{c}+\mm{jets}$ ($Wc+\mm{jets}$) cross sections are multiplied by an additional 1.5 (1.3). The $p_T$ distribution of the $W$ boson in MC simulation is reweighted to match the product of the $p_T$ distribution of the $Z$ boson measured in \dzero data \cite{zbosonpt} and the SM ratio of these two distributions, which was calculated at NLO using \resbos \cite{resbos}.

Other backgrounds include events from \zplus production, which include $Z$ bosons decaying to electron, muon or tau pairs. The LO \alpgen predictions are similarly corrected using the NLO calculation of \mcfm. The \zplus cross section is multiplied by 1.3, and the $Zc\bar{c}+\mm{jets}$ and $Zb\bar{b}+\mm{jets}$ cross sections by an additional 1.7 and 1.5, respectively. The simulated $p_T$ distribution of the $Z$ boson is reweighted to match the measured $p_T$ distribution in $Z \rightarrow \ell \ell$ \cite{zbosonpt}.

The single top quark background consists of $s$- and $t$-channel single top quark productions, which are normalized to the NLO cross sections of 1.04 and 2.26 pb \cite{singleTopXsec}, respectively. As the single top quark background yields only a few events passing all selection criteria described later, no effects are considered from the dependence of this background on $m_t$.

Diboson production ($WW$, $WZ$ and $ZZ$ bosons) processes are normalized to NLO cross sections, calculated with \mcfm, of 11.6 pb, 3.3 pb and 1.3 pb, respectively.\\

%**************************************************
%
%
%
%
%
\section{The \dzero Detector}
The \dzero detector \cite{d0detector} consists of several subdetectors designed for identification and 
reconstruction of the products of \ppbar collisions. A silicon microstrip tracker (SMT) \cite{smt,smt_l0} and central fiber tracker surround the interaction region for pseudorapidities\footnote{The pseudorapidity $\eta=-\ln\left[\tan(\theta/2)\right]$ is measured relative to the center of the detector, and $\theta$ is the polar angle with respect to the proton beam direction.} $|\eta| < 3$ and $|\eta| < 2.5$, respectively. These elements of the central tracking system are located within a superconducting solenoidal magnet generating a 1.9 T field, providing measurements for reconstructing event vertices and trajectories of charged particles. The SMT allows for a precision of $40~\mm{\mu m}$ or better for the reconstructed primary \ppbar interaction vertex (PV) in the plane transverse to the beam direction. The impact parameter of typical charged particle trajectories relative to the PV is determined with a precision between 20 and 50 $\mu$m depending on the number of SMT hits and particle momenta. The impact parameter and its measurement uncertainty are key components of lifetime-based identification of jets containing $b$ quarks \cite{bid_nim}. Particle energies are measured using a liquid argon sampling calorimeter that is segmented into a central calorimeter covering $|\eta| < 1.1$, and two end calorimeters extending the coverage to $|\eta| = 4.2$. Outside of the calorimetry, trajectories of muons are measured using three layers of tracking detectors and scintillation trigger counters, and an iron toroidal magnet generating a 1.8 T field between the first two layers. Plastic scintillator arrays are located in front of the end-calorimeter cryostats to measure the luminosity \cite{lumi_nim}.
%*****************************************************************
%
%

\section{Event Selection}
\label{toc:data_samples}
This analysis uses all the data recorded by the \dzero detector at $\sqrt{s}=1.96$ TeV. After applying data quality requirements, the data correspond to an integrated luminosity of 9.7 fb${}^{-1}$. The trigger selects \ljets events by requiring at least one lepton (electron or muon) with an efficiency of 95\% or 80\% for \ttbar events containing an electron or muon candidate, respectively.

\begin{figure*}[ht]
  \begin{center}
% ejets MVD
  \includegraphics[width=0.975\columnwidth,angle=0]{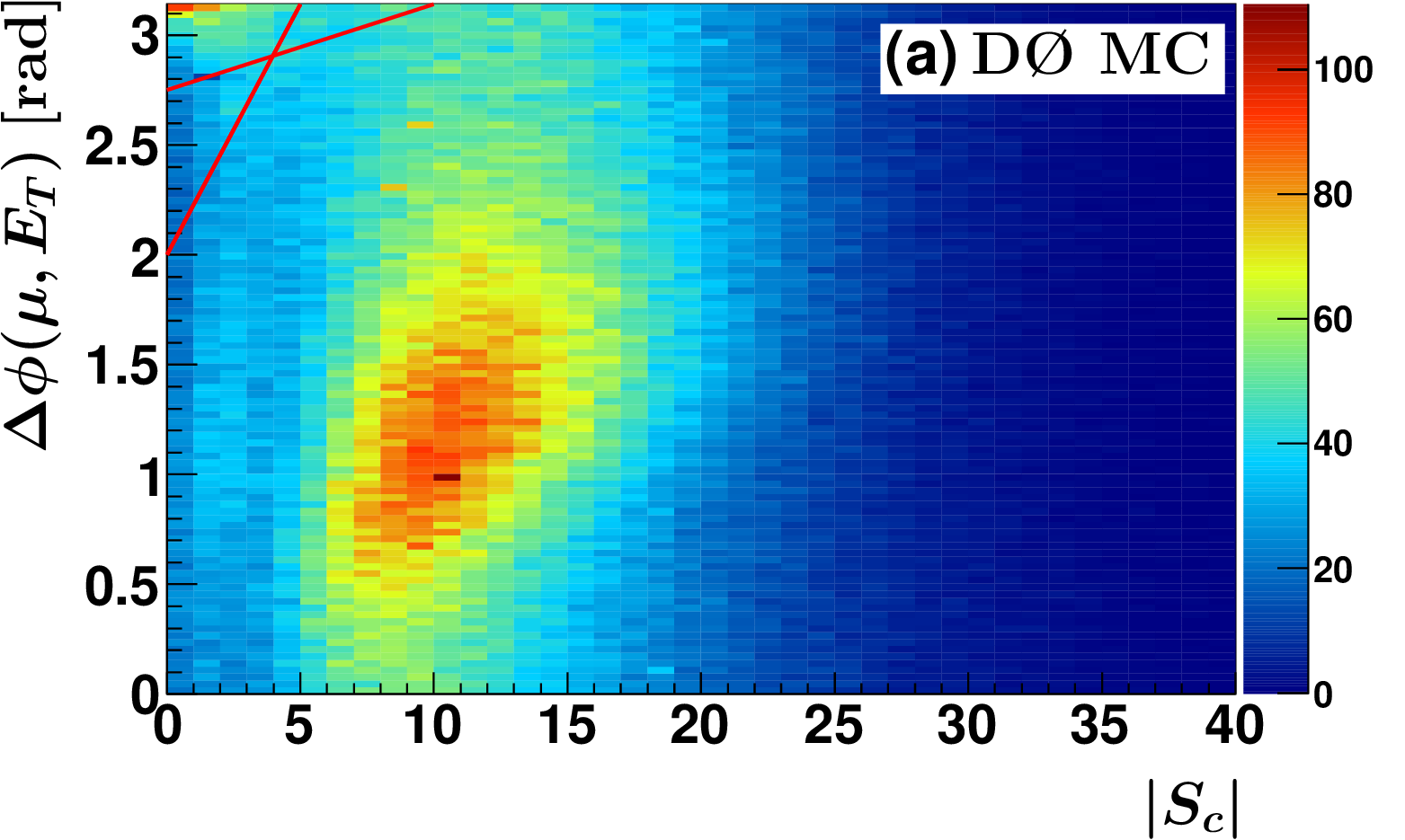}
% mujets MVD
  \includegraphics[width=0.975\columnwidth,angle=0]{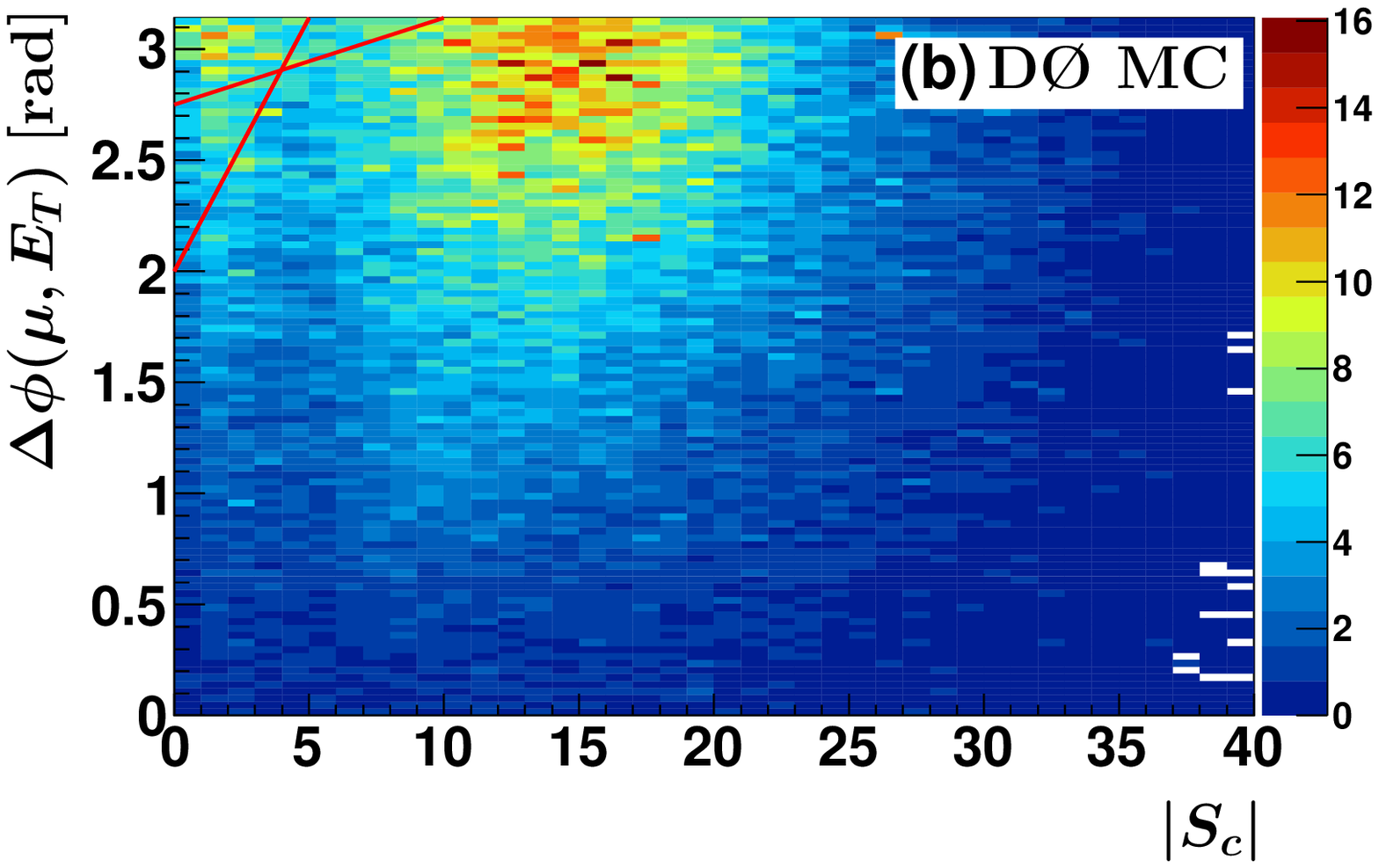}
  \end{center}
    \caption{The $|\cal{S}$%
$_c|$ versus $\Delta \phi(\mu, $ \metNo$)$ plane for (a) \ttbar events and (b) for \wplus background events. The selection requirements are indicated by the solid lines.} \label{fig:trkSig_sgbg}
\end{figure*}

Accepted events must have a reconstructed PV within $60~\mathrm{cm}$ of the center of the detector along the beam axis, one lepton with transverse momentum $p_{T}>20$~GeV and $\left|\eta\right|<1.1$ (for electrons) or $\left|\eta\right|<2$ (for muons), and $\met>20$~GeV. The measurement of \met is based on calorimetry. In addition, leptons are required to originate from the PV by demanding $|\Delta z(\ell, \mathrm{PV})| < 1~\mathrm{cm}$. A distance $\Delta R = \sqrt{\Delta\eta^2 + \Delta\phi^2}$ between a lepton and a jet of $\Delta R(\ell, \mathrm{closest~jet}) > 0.5$ is required to ensure that leptons are isolated \cite{muon_nim, elec_nim}. For the \muplus sample upper limits on the transverse mass of the reconstructed $W$ boson of $M_T^W < 250$ GeV and \met$<250$ GeV are applied to remove events in data with misreconstructed muon $p_T$. To further remove such events, we employ an additional requirement on the significance of the track curvature $\cal{S}$%
$_c$, which is defined as the ratio of the curvature, $\kappa$, and the expected uncertainty on $\kappa$ measured for the track associated with the muon. We employ two selection requirements with different slopes in the azimuthal ($\Delta \phi$) vs $\cal{S}$%
$_c$ plane: \mbox{$(-70 + 25.47 \cdot \Delta \phi(\mu, $ \metNo$)) < |\cal{S}$%
$_c|$} and \mbox{$(-8.76 + 4.38 \cdot \Delta \phi(\mu, $ \metNo$)) < |\cal{S}$%
$_c|$}. Figure \ref{fig:trkSig_sgbg}(a) shows these requirements indicated by the solid lines in the $|\cal{S}$%
$_c|$ versus $\Delta \phi(\mu, $ \metNo$)$ plane for \ttbar events and \ref{fig:trkSig_sgbg}(b) for \wplus background events. The cut on $\cal{S}$%
$_c$ removes low momentum muons misreconstructed at high momenta while keeping 97\% of the leptons stemming from \ttbar decays.
A minimum separation in azimuth of \mbox{$\Delta \phi(\ell, $ \metNo$) > 0.5$} is imposed between the direction of the lepton and the direction of the missing momentum, to reduce multijet background caused by the misidentification of a jet as a lepton and the consequent impact on the accompanying \met. Further reduction of the multijet background is achieved by requiring an additional minimum separation in azimuth between the isolated lepton and \met: \mbox{$\Delta \phi(e, $ \metNo$) > 2.2 - 0.045 \cdot \metNo/\mm{GeV}$} and \mbox{$\Delta \phi(\mu, $ \metNo$) > 2.1 - 0.035 \cdot \metNo/\mm{GeV}$}. After correcting the energy of the jet to the particle level \cite{jescorrection} at least four jets with $p_{T}>20$~GeV and $\left|\eta\right|<2.5$ are required. The jet with highest $p_T$ is also required to have $p_{T}>40$~GeV.

Because of the high instantaneous luminosity provided by the Tevatron, additional \ppbar collisions may occur within the same bunch crossing. As noted above, events from randomly selected beam crossings with the same instantaneous luminosity are overlaid on the simulated events, which are reweighted to match the luminosity profile observed in data. To suppress jets from these additional collisions, jets are required to contain two tracks consistent with originating from the PV. At least one of the jets must be selected as likely to originate from a $b$ quark ($b$ tagged) using a multivariate discriminant (MVD) \cite{bid_nim}. The discriminant combines variables that characterize the presence and properties of secondary vertices and tracks within jets. The MVD identification of jets containing $b$ quarks has an efficiency of approximately $60\%$, with a light quark misidentification rate of approximately 1.2\%. Events containing more than one isolated muon or electron, which satisfy the lepton requirements discussed above, are rejected.\\

\section{Sample Composition}
\label{toc:sampleCompSec}
\begin{figure*}[ht]
  \begin{center}
% ejets MVD
  \includegraphics[width=0.975\columnwidth,angle=0]{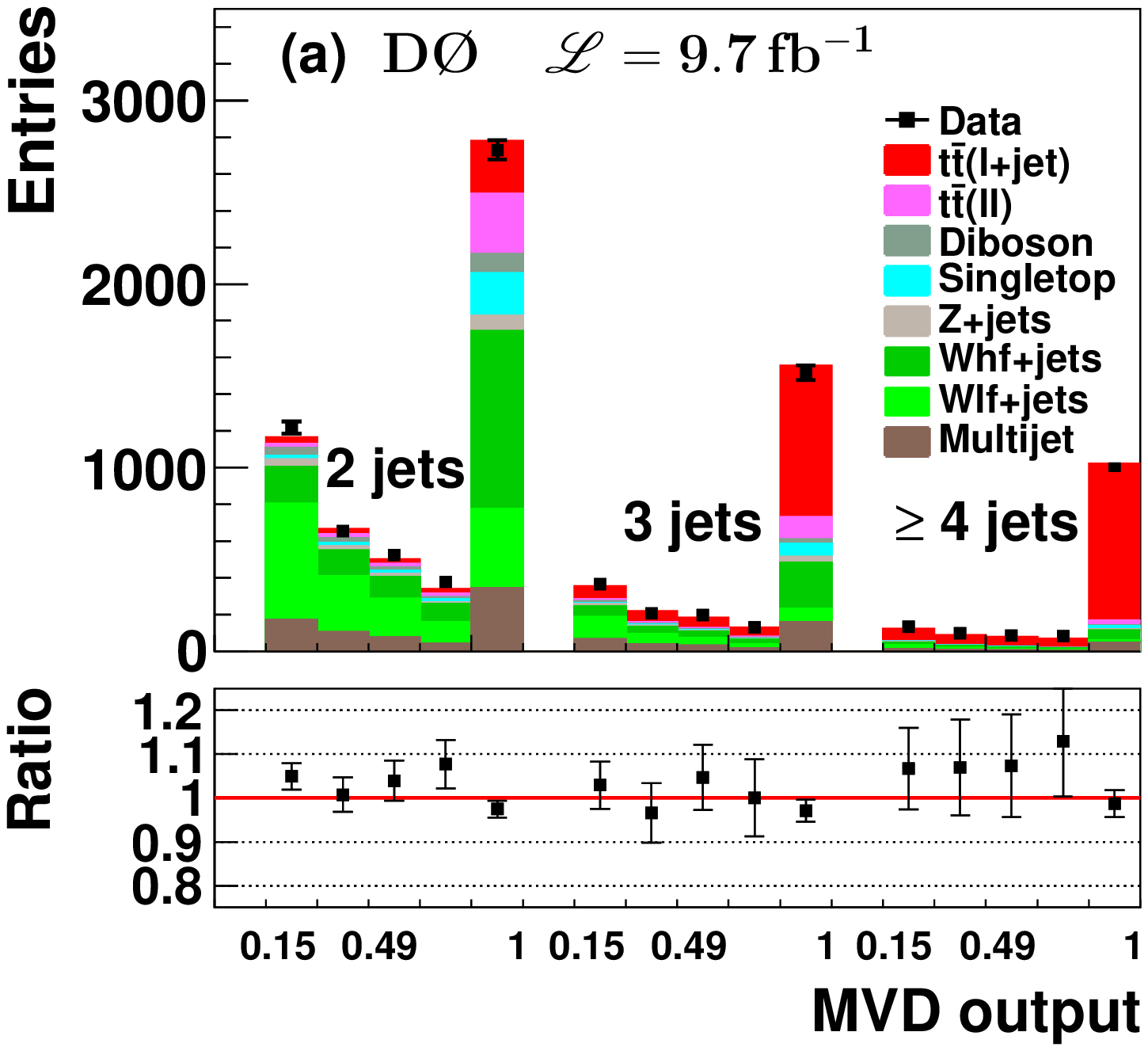} \hspace{10pt}
% mujets MVD
  \includegraphics[width=0.975\columnwidth,angle=0]{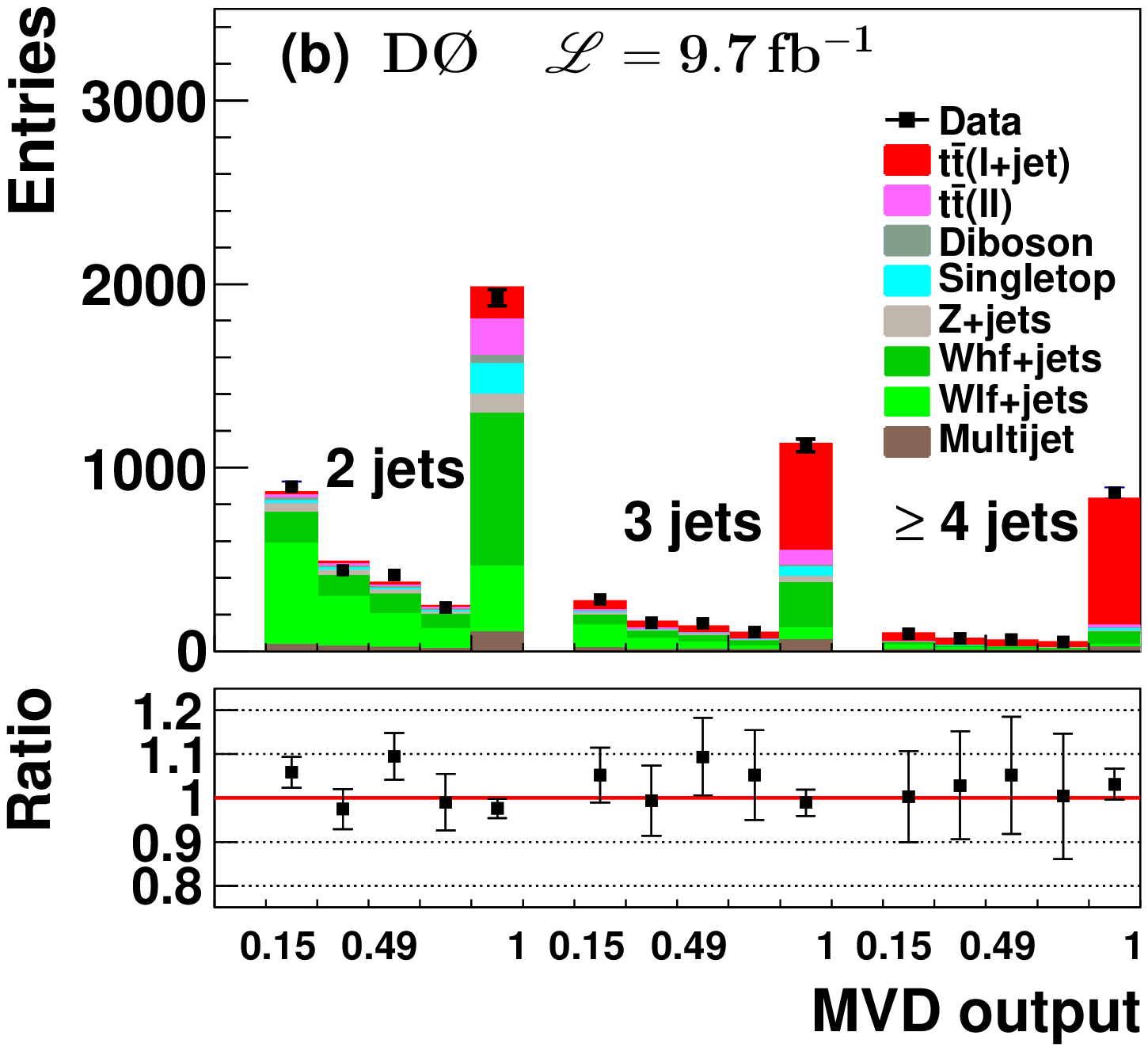}
  \end{center}
    \caption{Distributions of the MVD $b$ identification output distribution for the \lplustw, \lplusth and \lplusgefo data sample in the (a) \eplus and (b) \muplus decay channel. The data are compared to the sum of predicted contributions from signal and background processes. More details on how the sample composition is derived can be found in the text.} \label{fig:sampleComp}
\end{figure*}

Background contributions are categorized into instrumental background and irreducible background from processes with final states similar to \ttbar. %Instrumental background arises from \ttbar events where both $W$ bosons decay leptonically, but only one of the leptons is identified or is within the defined acceptance. 
Instrumental background is due to multijet processes where a jet is misidentified as an electron in the \eplus channel, or when a muon originating from the semileptonic decay of a heavy hadron appears to be isolated in the \muplus channel. Data-driven \cite{matrixMethod, Publ54_xsec} and MC simulation methods are employed to model the instrumental background. The irreducible background processes are estimated using MC simulations described in Sec.\ \ref{toc:generators}. Most of this background arises from \wplus production, and to constrain it we use the \lplustw and \lplusth data (dominated by \wplus production) in addition to the \lplusgefo sample (dominated by \ttbar production). We determine the sample composition from a simultaneous fit for the \ttbar cross section and the heavy-flavor contribution originating from \wplus. The fit is made to the MVD $b$ identification output distribution; Fig.\ \ref{fig:sampleComp} shows the distribution after applying the fit results for the \lplustw, \lplusth and \lplusgefo data sample in the \ref{fig:sampleComp}(a) \eplus and \ref{fig:sampleComp}(b) \muplus decay channel.
The simultaneous fit yields a \wplus heavy-flavor scale factor $s_{\mm{fit}}^{\mm{WHF}} = 0.89 \pm 0.08$ to be applied to the $Wb\bar{b}+\mm{jets}$ and $Wc\bar{c}+\mm{jets}$ contributions in addition to the factors discussed in Sec.\ \ref{toc:generators}. Similar procedures were used in previous measurements by \dzero \cite{Publ54_xsec}. The simultaneous fit to the \lplustw, \lplusth and \lplusgefo samples yields a \ttbar cross section of $\sigma_{\mm{fit}}^{t\bar{t}} = 8.00 \pm 0.40\thinspace(\mm{stat.})$ pb. We verified that there is no need for an additional scale factor to accommodate the \zplus heavy-flavor contributions $s_{\mm{fit}}^{\mm{ZHF}}$ by using a modified version of the simultaneous fit taking into account $s_{\mm{fit}}^{\mm{ZHF}}$ instead of $s_{\mm{fit}}^{\mm{WHF}}$. The $\sigma_{\mm{fit}}^{t\bar{t}}$ serves as an initial value of the \ttbar cross section in the \ttbar differential cross section measurement using inclusive four-jet data.\\

The total inclusive \ttbar cross section is also calculated using only events with at least four jets from the three differential distributions by integrating all bins of each of the cross section distributions, as presented below in Sec. \ref{toc:xsec} and average the resulting three inclusive cross sections as discussed in Sec. \ref{toc:results}. This yields a compatible value of $\sigma(p\bar{p} \rightarrow t\bar{t}) =  8.0 \pm 0.7\thinspace(\mm{stat.}) \pm 0.8\thinspace(\mm{syst.})~\mm{pb}$. The \ttbar contributions in the following plots are derived employing \mcatnlo simulated events normalized to this measured inclusive $t\bar{t}$ cross section of $8.0~\mathrm{pb}$.

\begin{figure*}[ht]
  \begin{center}
% NJets
  \includegraphics[width=0.975\columnwidth,angle=0]{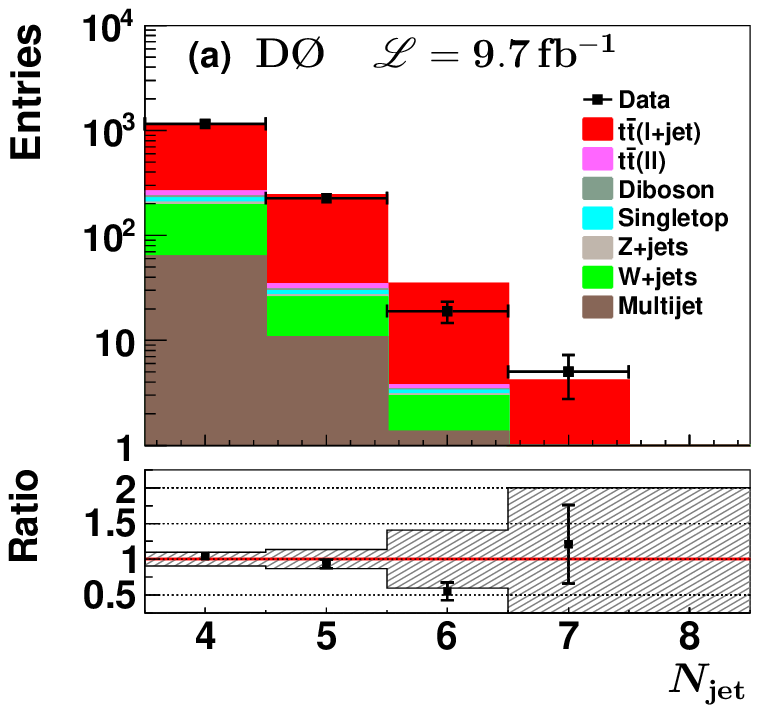}
% Ht
  \includegraphics[width=0.975\columnwidth,angle=0]{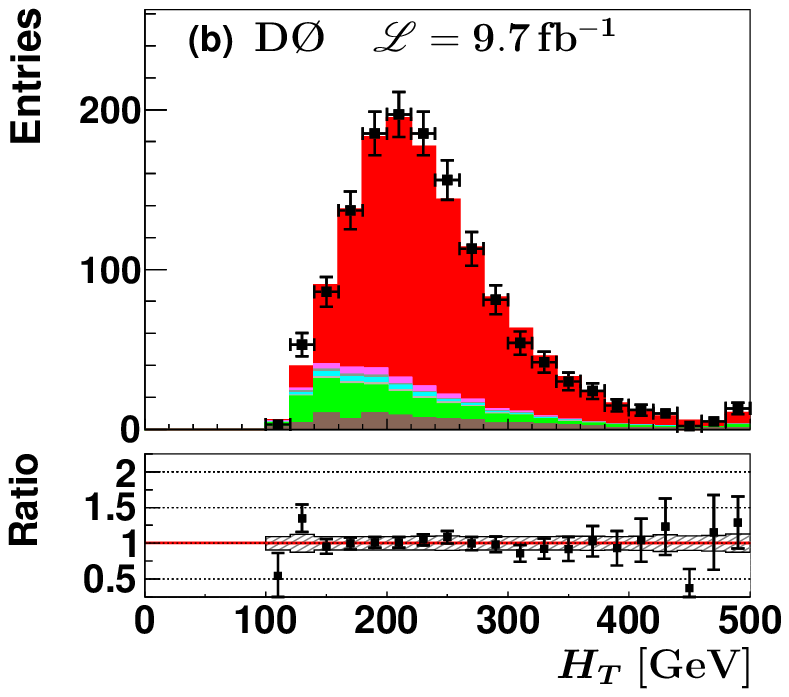}
% met
  \includegraphics[width=0.975\columnwidth,angle=0]{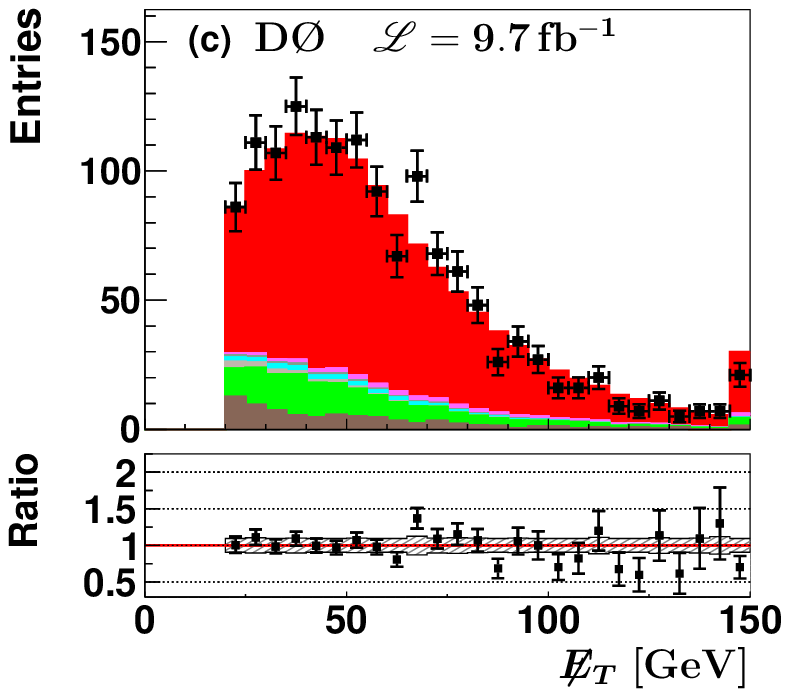}
% lep pt
  \includegraphics[width=0.975\columnwidth,angle=0]{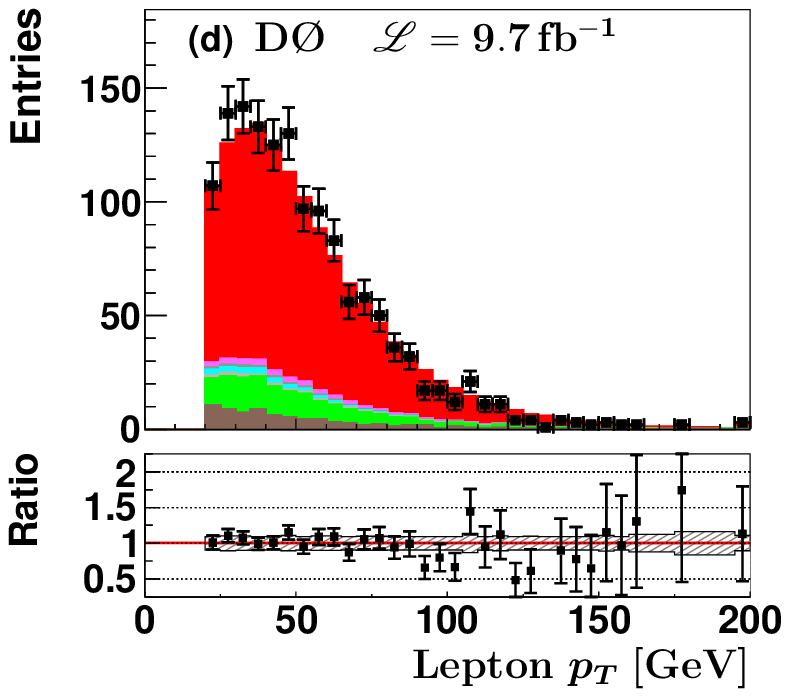}
  \end{center}
    \caption{Distributions of (a) the number of jets, (b) the scalar sum of the $p_T$ values of the lepton and jets, (c) $\met$, and (d) lepton $p_T$ for the \eplus final state. The data are compared to the sum of predicted contributions from signal and background processes. The signal contribution is derived employing \mcatnlo simulated events normalized to the measured inclusive $t\bar{t}$ cross section of $8.0~\mathrm{pb}$. The highest bin in the histograms is used as an overflow bin. The ratios of data to the sum of the signal and all background contributions are shown in the panels below the distributions. The bands show the $1$ s.\!\,d.\ combined systematic uncertainties on the sum of the signal and background contributions.} \label{fig:njetsall_ej}
\end{figure*}

\begin{figure*}[ht]
  \begin{center}
% NJets
  \includegraphics[width=0.975\columnwidth,angle=0]{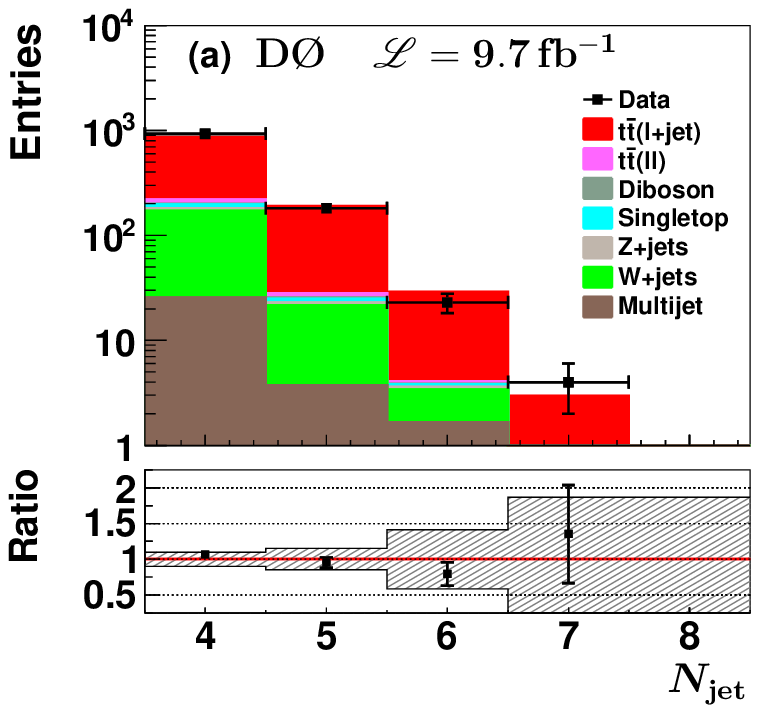}
% Ht
  \includegraphics[width=0.975\columnwidth,angle=0]{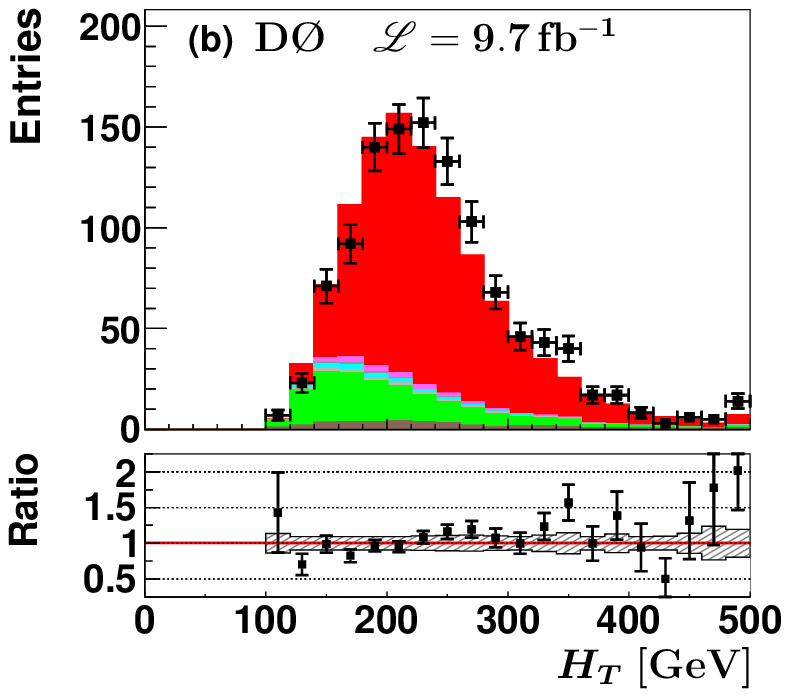}
% met
  \includegraphics[width=0.975\columnwidth,angle=0]{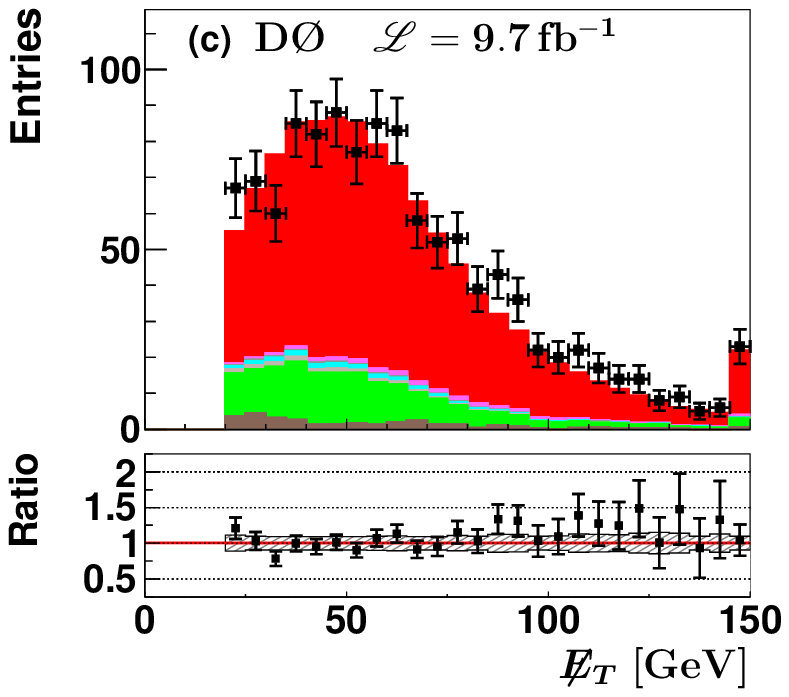}
% lep pt
  \includegraphics[width=0.975\columnwidth,angle=0]{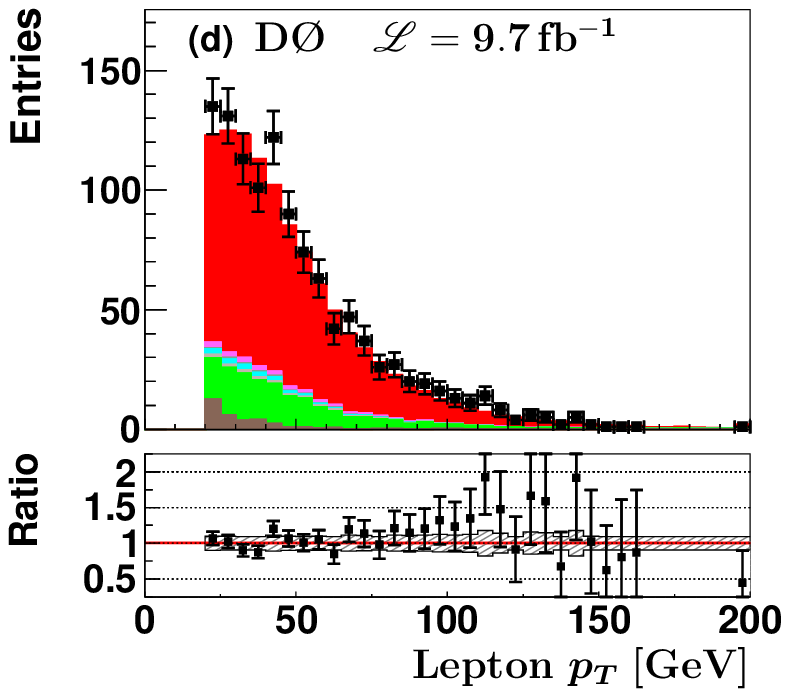}
  \end{center}
    \caption{Distributions of (a) the number of jets, (b) the scalar sum of the $p_T$ values of the lepton and jets, (c) $\met$, and (d) lepton $p_T$ for the \muplus final state. The data are compared to the sum of predicted contributions from signal and background processes. The signal contribution is derived employing \mcatnlo simulated events normalized to the measured inclusive $t\bar{t}$ cross section of $8.0~\mathrm{pb}$. The highest bin in the histograms is used as an overflow bin. The ratios of data to the sum of the signal and all background contributions are shown in the panels below the distributions. The bands show the $1$ s.\!\,d.\ combined systematic uncertainties on the sum of the signal and background contributions.} \label{fig:njetsall_mu}
\end{figure*}
Figures \ref{fig:njetsall_ej} and \ref{fig:njetsall_mu} demonstrate, respectively, the quality of the modeling of the selected events in the \eplus and \muplus sample with the background and signal contributions. The signal contribution is derived employing \mcatnlo simulated events normalized to the measured inclusive $t\bar{t}$ cross section of $8.0~\mathrm{pb}$. %The last bin in the histograms is used as an overflow bin. 
The expected composition of the sample after the final selection is given in Table \ref{tab:yields_runIIab_1t_emujets}.

\begin{table}[htdp]
    \caption{Expected number of events with at least four jets due to each process (uncertainties are statistical and systematical added in quadrature). The sample composition is determined as discussed in Sec.\ \ref{toc:sampleCompSec}. Events in the \ttbar dilepton decay channel are denoted by $\ell \ell$.} \label{tab:yields_runIIab_1t_emujets}
\begin{center}
\begin{ruledtabular}
\begin{tabular}{l l l}
\multicolumn{1}{c}{Process} & \multicolumn{1}{c}{\muplus} & \multicolumn{1}{c}{\eplus} \\ \hline
% Multijet &          $\hphantom{00} 31.1 \pm 10.0$  & $\hphantom{00} 75.1 \pm 13.0$ \\
% \wjets	  &         $\hphantom{0} 164.9 \pm \hphantom{0} 3.1$ & $\hphantom{0} 148.8 \pm \hphantom{0} 2.6$ \\
% Diboson   &         $\hphantom{000} 9.1 \pm \hphantom{0} 0.3$ & $\hphantom{00} 10.5 \pm \hphantom{0} 0.3$ \\
% \zjets    &         $\hphantom{00} 11.9 \pm \hphantom{0} 0.4$ & $\hphantom{00} 12.4 \pm \hphantom{0} 0.4$ \\
% Single top &        $\hphantom{00} 16.1 \pm \hphantom{0} 0.2$ & $\hphantom{00} 21.8 \pm \hphantom{0} 0.3$ \\
% \ttbar, $\ell \ell$&$\hphantom{00} 22.6 \pm \hphantom{0} 0.2$ & $\hphantom{00} 33.5 \pm \hphantom{0} 0.3$ \\ \hline
%  $\sum\,\mm{bgs}$ & $\hphantom{0} 254.4 \pm 10.5$ & $\hphantom{0} 302.1 \pm 13.3$ \\
% % using 8.27, the measured one
% \ttbar, \ljets    & $\hphantom{0} 838.7 \pm \hphantom{0} 3.2$  & $1088.7 \pm \hphantom{0} 3.8$ \\ \hline 
%  $\sum\,\mm{(sig + bgs)}$ & $1093.1$ $\pm$ $11.0$ & $1390.8 \pm 13.8$ \\

Multijet &          $\hphantom{00} 31.1 \pm 10.0$  & $\hphantom{00} 75.1 \pm \hphantom{0} 56.3$ \\
\wjets	  &         $\hphantom{0} 164.9 \pm 15.9$ & $\hphantom{0} 148.8 \pm \hphantom{0} 14.3$ \\
Diboson   &         $\hphantom{000} 9.1 \pm \hphantom{0} 0.8$ & $\hphantom{00} 10.5 \pm \hphantom{00} 0.9$ \\
\zjets    &         $\hphantom{00} 11.9 \pm \hphantom{0} 1.2$ & $\hphantom{00} 12.4 \pm \hphantom{00} 1.5$ \\
Single top &        $\hphantom{00} 16.1 \pm \hphantom{0} 2.2$ & $\hphantom{00} 21.8 \pm \hphantom{00} 3.0$ \\
\ttbar, $\ell \ell$&$\hphantom{00} 22.6 \pm \hphantom{0} 2.0$ & $\hphantom{00} 33.5 \pm \hphantom{00} 2.9$ \\ \hline
 $\sum\,\mm{bgs}$ & $\hphantom{0} 254.4 \pm 19.1$ & $\hphantom{0} 302.1 \pm \hphantom{0} 58.3$ \\
% % using 8.27, the measured one
\ttbar, \ljets    & $\hphantom{0} 838.7 \pm 72.5$  & $1088.7 \pm \hphantom{0} 94.2$ \\ \hline 
 $\sum\,\mm{(sig + bgs)}$ & $1093.1 \pm 75.0$ & $1390.8 \pm 110.8$ \\
 \multicolumn{1}{l}{Data} & \multicolumn{1}{c}{1137} & \multicolumn{1}{c}{1403} \\
 \end{tabular}
\end{ruledtabular}
 \end{center}
\end{table}

%***********************************************************************
%
%
%
%
\section{Extraction of the Signal}
\label{toc:signalExtraction}
\begin{figure*}[ht]
  \begin{center}
   \includegraphics[width=0.975\columnwidth,angle=0]{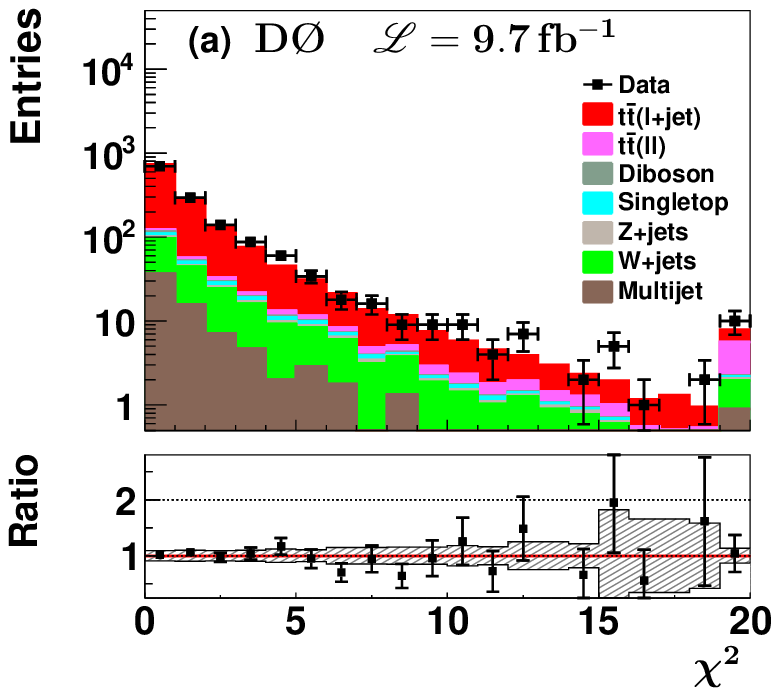}
   \includegraphics[width=0.975\columnwidth,angle=0]{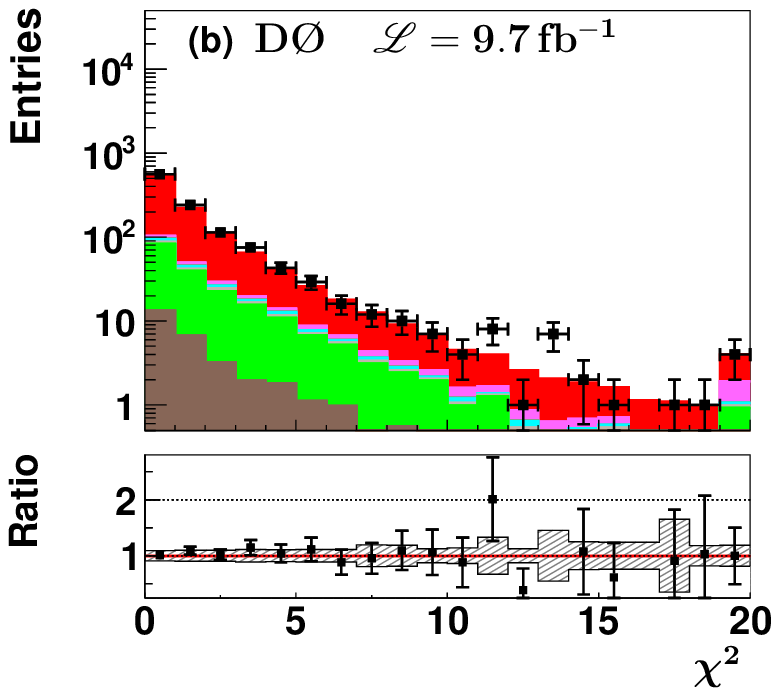}
  \end{center}
    \caption{Distribution of $\chi^2$ for the best solution with lowest $\chi^2$ for the (a) \eplus and (b) \muplus final states. The data are compared to the sum of predicted contributions from signal and background processes. The signal contribution is derived employing \mcatnlo simulated events normalized to the measured inclusive $t\bar{t}$ cross section of $8.0~\mathrm{pb}$. The highest bin in the histograms is used as an overflow bin. The ratios of data to the sum of the signal and all background contributions are shown in the panels below the distributions. The bands show the $1$ s.\!\,d.\ combined systematic uncertainties on the sum of the signal and background contributions.} \label{fig:hitfitChi2_ejets}
\end{figure*}

\begin{figure*}[ht]
  \begin{center}
    \includegraphics[width=0.975\columnwidth,angle=0]{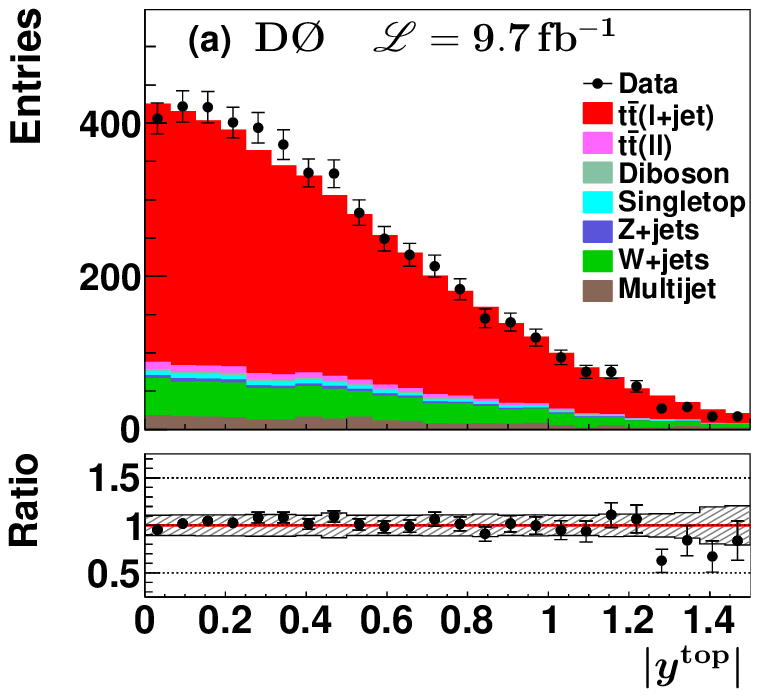}
    \includegraphics[width=0.975\columnwidth,angle=0]{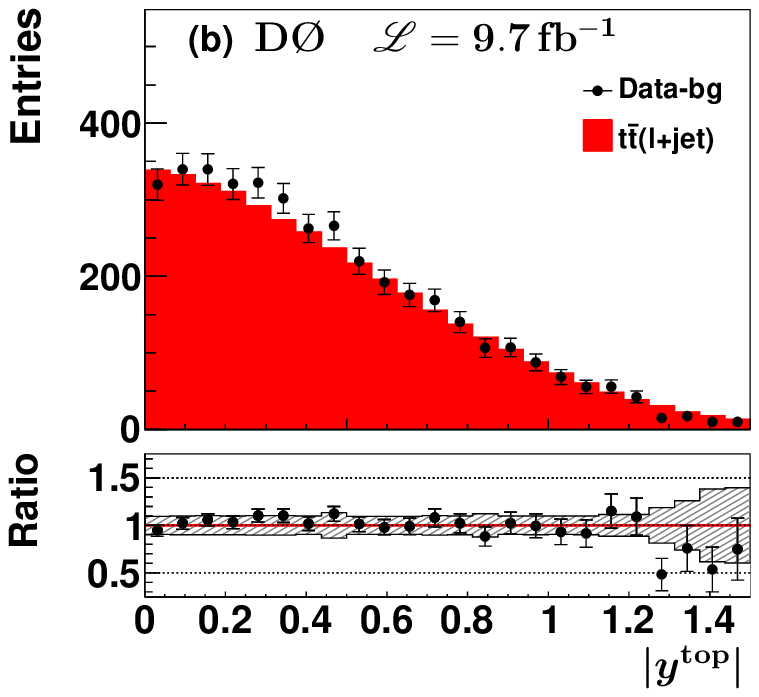}
  \end{center}
    \caption{Distribution of \aetat, (a) compared to the sum of predicted contributions from signal and background processes, and (b) the background-subtracted distribution. The signal contribution is derived employing \mcatnlo simulated events normalized to the measured inclusive $t\bar{t}$ cross section of $8.0~\mathrm{pb}$. The lower panels indicate the ratio of the data to (a) the sum of the signal and all background processes, and (b) to the signal process only.}\label{fig:aetatcontrolBg}
\end{figure*}

\begin{figure*}[ht]
  \begin{center}
    \includegraphics[width=0.975\columnwidth,angle=0]{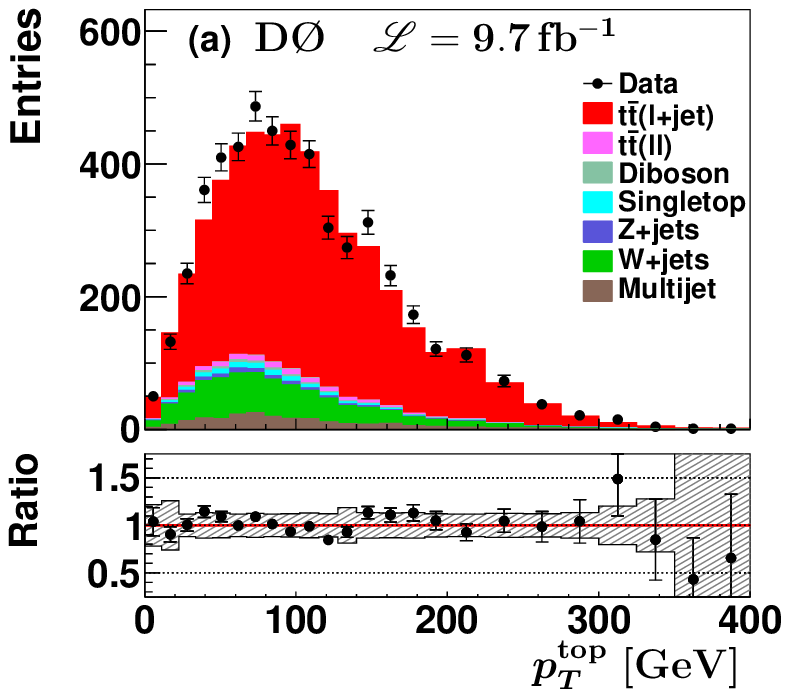}
    \includegraphics[width=0.975\columnwidth,angle=0]{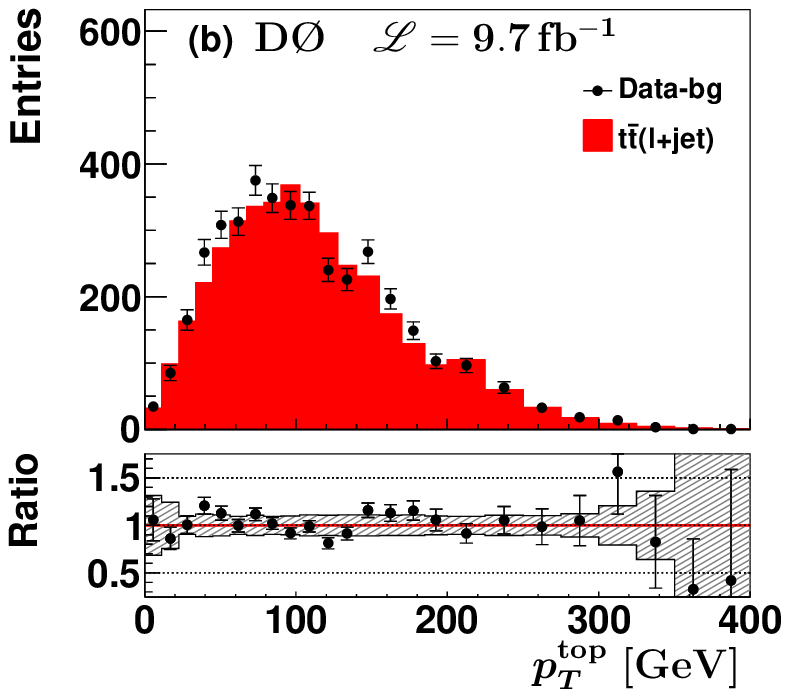}
  \end{center}
    \caption{Distribution of \ptt, (a) compared to the sum of predicted contributions from signal and background processes, and (b) the background-subtracted distribution. The signal contribution is derived employing \mcatnlo simulated events normalized to the measured inclusive $t\bar{t}$ cross section of $8.0~\mathrm{pb}$. The lower panels indicate the ratio of the data to (a) the sum of the signal and all background processes, and (b) to the signal process only.}\label{fig:pttcontrolBg}
\end{figure*}

To reconstruct the four-vectors of the full \ttbar decay chain, $t\bar{t} \rightarrow W^+b+W^-\bar{b} \rightarrow (q\bar{q}')b+(\ell \nu)\bar{b}$, we use a constrained kinematic reconstruction algorithm \cite{d0_hitfit} that takes into account experimental resolutions. In total the algorithm uses 18 parameters based on the measurements of jets, leptons and \met. The masses of the $W$ boson and the $t$ quark are fixed to $80.4~\mathrm{GeV}$ and $172.5~\mathrm{GeV}$, respectively. The \met provides the initial estimate for the $p_T$ of the neutrino. The longitudinal momentum $p_z(\nu)$ is estimated by constraining the mass of the $W$ boson decay products to $80.4~\mathrm{GeV}$. This yields a quadratic equation in $p_z(\nu)$ with two solutions. These solutions, together with the $12$ possible jet-quark assignments yield 24 possible solutions to the kinematic reconstruction algorithm. The large number of solutions is reduced by assigning $b$-tagged jets to $b$ quarks. The solution with the best $\chi^2$ for assigning the reconstructed objects to the parton-level quantities serves as the input to the unfolding (see Sec.\ \ref{toc:unfolding_intro}). This solution corresponds to the correct assignment of the jets to the quarks from the \ttbar decay in MC events in 80\% of the cases. The observed and expected distributions in $\chi^2$ are compared in Fig.\ \ref{fig:hitfitChi2_ejets}.\\
The modeling of signal and background processes is verified through a comparison of the data to the number of expected \ttbar signal events and the sum of all background contributions. The expected \ttbar contribution is derived employing \mcatnlo simulated events normalized to the measured inclusive cross section of $8.0~\mathrm{pb}$. 
\begin{figure*}[ht]
  \begin{center}
    \includegraphics[width=0.975\columnwidth,angle=0]{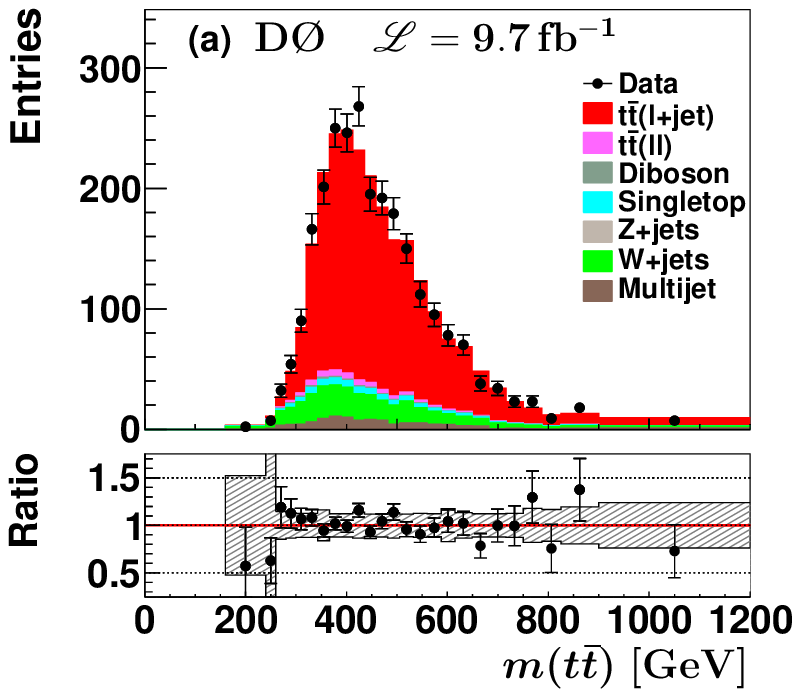}
    \includegraphics[width=0.975\columnwidth,angle=0]{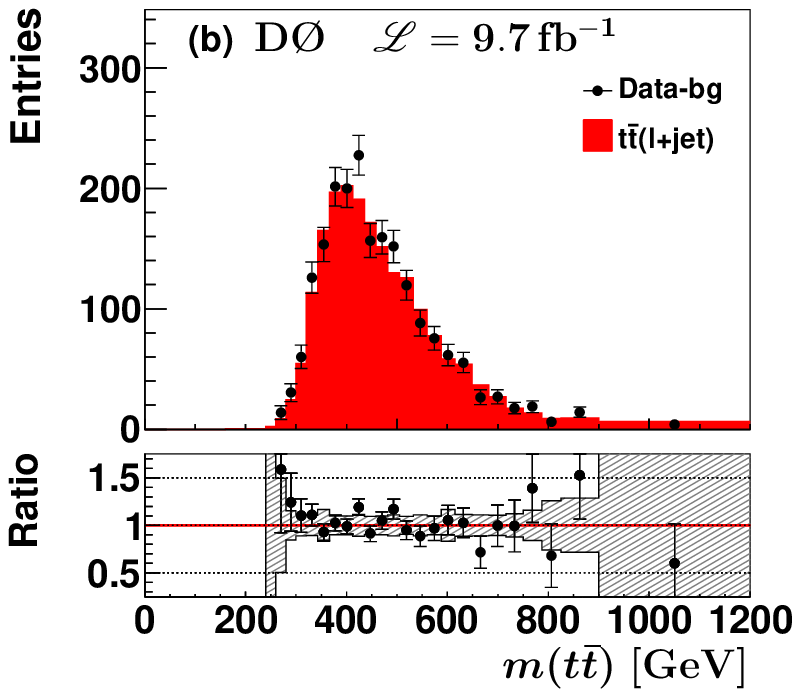}
  \end{center}
    \caption{Distribution of \mTT, (a) compared to the sum of predicted contributions from signal and background processes, and (b) the background-subtracted distribution. The signal contribution is derived employing \mcatnlo simulated events normalized to the measured inclusive $t\bar{t}$ cross section of $8.0~\mathrm{pb}$. The lower panels indicate the ratio of the data to (a) the sum of the signal and all background processes, and (b) to the signal process only.}\label{fig:mTTcontrolBg}
\end{figure*}
Figures \ref{fig:mTTcontrolBg}--\ref{fig:pttcontrolBg} show the reconstructed \mTT, \aetat, and \ptt distributions before unfolding. The \aetat and \ptt distributions include both $W \rightarrow \ell \nu$ and $W \rightarrow q\bar{q}'$ decay modes (two entries per event). The resolutions in the two decay modes are similar; hence they are combined. The distributions in (a) of Figs.\ \ref{fig:mTTcontrolBg}--\ref{fig:pttcontrolBg} show the data compared to the \ttbar signal and background processes, while (b) shows the background-subtracted data. The data and its description by the sum of signal and background processes agree within uncertainties.

%********************************************************************
%
%
%
%
%
\section{Measurement technique}
\label{toc:unfolding_intro}
Measurements involving top quarks benefit from the very short lifetime of the $t$ quark, since it decays before it can hadronize. Effects of hadronization and QCD corrections are thus reduced. Moreover, at Tevatron energies the transverse momentum of \ttbar pairs is almost always smaller than \mTT and production is central, so that almost the entire phase space of \ttbar production is within the detector acceptance. Corrections to measured quantities as well as their uncertainties are therefore small, leading to well measured top-quark cross sections.\\

The differential cross sections are defined for parton-level top quarks including off-shell effects and are corrected for detector and QCD effects using a regularized matrix unfolding procedure \cite{unfold1, unfold2}. This procedure reduces the influence of model dependencies in the cross section determination and introduces correlations among the bins used in the measurement. These correlations are minimized by regularization. Unfolding event migrations relies on a migration matrix ($A$), which describes the relation between the generated distribution of a variable ($\vec{x}_{\mathrm{gen}}$) and its reconstructed distribution ($\vec{y}_{\mathrm{rec}}$) as $A \vec{x}_{\mathrm{gen}} = \vec{y}_{\mathrm{rec}}$. Each matrix element $A_{ij}$ is the probability for an event originating from bin $j$ of $\vec{x}_{\mathrm{gen}}$ to be measured in bin $i$ of $\vec{y}_{\mathrm{rec}}$. The migration matrix is based on the simulation of the \dzero detector. The reconstruction-level bins used in the migration matrix are twice as narrow as the generator level bins, in order to provide detailed information on the bin-to-bin migrations, and improve the accuracy of the unfolding \cite{blobel}. The generated distribution $\vec{x}_{\mathrm{gen}}$ can be estimated using $A^{\dagger}$, the pseudoinverse \cite{pseudoinverse} of the matrix $A$: $\vec{x}_{\mathrm{gen}} = A^{\dagger} \vec{y}_{\mathrm{rec}}$. As with ordinary matrix inversion, this results in large contributions that lack statistical significance. Such contributions can be minimized by imposing regularization, which leads to an effective cutoff of the insignificant terms. We employ regularized unfolding as implemented in the \tunfold package \cite{tunfold}. The regularization is based on the derivative of the distribution and is done in twice as many bins as are used in the final results. An insufficient regularization admits fluctuations into the unfolded result, whereas excessive regularization overly biases the measurement toward the MC generated distribution. The value of the regularization strength is determined using the so-called L-curve approach \cite{tunfold} that balances the consistency of the unfolded data $x$ with the initial data $y$ against the scatter of $x$. The scatter of $x$ can be caused by fluctuations in cases in which an insufficient regularization is chosen. A $\chi^2$ statistic measures the tension between $x$, the data and the scatter of $x$. Within the earlier mentioned bounds, a systematic uncertainty is derived for this procedure as discussed in Sec.\ \ref{toc:xsec_sys_reg}. The statistical uncertainties of the differential measurements are computed analytically with \tunfold and verified using an ensemble of simulated pseudo--data sets. The covariance matrix is calculated by propagating the uncertainties of the reconstructed distribution $\vec{y}_{\mathrm{rec}}$ through the unfolding process.

%***************************************
%
%
%
\section{Cross Section Determination}
\label{toc:xsec}
Equation (\ref{eqn:xsecDef}) is used to calculate the differential \ttbar cross section $\sigma_i$ as a function of
the observable $X$, where $i$ denotes an individual bin, and $\Delta X_i$ its width.\\
\begin{equation}
 \label{eqn:xsecDef}
\dfrac{d\sigma_i}{dX} = \dfrac{N^{\mm{unfold}}_i}{{\mathscr{L}} \cdot B \cdot \Delta X_i}~.
\end{equation}
The unfolded number of signal events $N^{\mm{unfold}}_{i}$ is corrected for the branching fraction $B$ into the \ljets decay channel of $0.342 \pm 0.02$ \cite{pdg} and used to obtain the cross section for the total integrated luminosity $\mathscr{L}$ that corresponds to the selection requirements, including data quality cuts. The branching fraction used in Eq.\ (\ref{eqn:xsecDef}) includes electrons and muons originating from the decay of $\tau$ leptons. The number of expected background events is estimated through MC simulations and data-driven methods and is subtracted from data to determine $N^{\mm{unfold}}_{i}$. The numbers of background-subtracted events are corrected for effects due to limited detector resolution and efficiency by means of the regularized matrix unfolding as discussed in Sec.\ \ref{toc:unfolding_intro}. By using this procedure, the data are corrected for all detector effects including those from trigger, selection and $b$-tagging efficiencies and for the kinematic and geometric acceptance.

\section{Systematic Uncertainties}
\label{toc:xsec_sys}
Systematic uncertainties are assessed by varying the values of a specific parameter used in the modeling of the data, and repeating the analysis. Unless otherwise stated, the magnitude of the parameter modifications is obtained from alternative calibrations of the MC simulation. The migration matrix and the background contributions are extracted from these different MC models, while the regularization strength is fixed to that for the nominal unfolded data. The difference between the nominal unfolded data and unfolded data, including a modification due to a specific parameter serves as the estimate of an individual source of systematic uncertainty. Individual sources of systematic uncertainty are added in quadrature for each bin of a differential cross section. The largest uncertainties usually arise at large values of \mTT, \aetat, or \ptt, where there are fewer events. Table \ref{tab:syst_uncorr} summarizes the systematic uncertainties on the inclusive and differential cross sections. Numbers stated in the column denoted with $|\delta_{\mm{diff}}|$ illustrate the size of the systematic uncertainties in individual bins of the differential measurements.

\begin{table}[tp]%
\caption {\label{tab:syst_uncorr} Sources of systematic uncertainties. The uncertainty from each source on the inclusive cross section is given in the second column. Systematic uncertainties in the binned values of the differential cross sections vary within the range given in the last column.}
\centering %
\begin{ruledtabular}
\begin {tabular}{lcc}
\multicolumn{1}{l}{Source of uncertainty} & \multicolumn{2}{c}{Uncertainties, \%} \\
\multicolumn{1}{l}{} & \multicolumn{1}{c}{$\delta_{\mm{incl}}$}
                     & \multicolumn{1}{c}{$|\delta_{\mm{diff}}|$} \\ \hline
% Signal modeling & 				$^{+5.2}_{-4.4}$ &  $4.0$ - $14.2$ \T \\
% PDF & 					$^{+3.0}_{-3.4}$ &  $0.9$ - $\hphantom{0} 4.4$  \T \\
% Detector Modeling & 				$^{+4.0}_{-4.1}$ &  $3.1$ - $13.7$ \T \\
Signal modeling & 				$+5.2$/$-4.4$ &  $4.0$ -- $14.2$ \T \\
PDF & 						$+3.0$/$-3.4$ &  $0.9$ -- $\hphantom{0} 4.4$  \T \\
Detector modeling & 				$+4.0$/$-4.1$ &  $3.1$ -- $13.7$ \T \\
Sample composition & 				$\!\pm 1.8$       &  $2.8$ -- $\hphantom{0} 9.2$  \T \\
Regularization strength & 			$\!\pm 0.2$       &  $0.8$ -- $\hphantom{0} 2.1$  \T \\
Integrated luminosity & 			$\!\pm 6.1$       &  $6.1$ -- $\hphantom{0} 6.1$ \T \\ \hline
% Total systematic uncertainty 	& 		$^{+9.6}_{-9.3}$ &  $8.5$ - $23.1$ \T \\
Total systematic uncertainty 	& 		$+9.6$/$-9.3$ &  $8.5$ -- $23.1$ \T \\
\end {tabular}
\end{ruledtabular}
\end {table}

\subsection{Modeling of signal}
The effect of NLO corrections on the matrix element for \ttbar production is estimated by comparing \ttbar events generated with \mcherwig to those from \alppyt. From a comparison of \alppyt to \alpher, we find that the effects of hadronization uncertainties are less than those from the inclusion of higher-order effects. The top mass is varied within its uncertainty of $\pm 1$ GeV \cite{TevatronMassCombo}. An additional uncertainty on the signal arises from the relatively poor modeling of the reconstructed transverse momentum of the \ttbar pair \ptTT at \dzero \cite{d0_afb}. A systematic uncertainty is estimated by reweighting the distribution of the reconstructed \ptTT in the MC simulation to the one observed in \dzero data.

\subsection{Parton distributions functions} 
The uncertainty on the cross sections due to the uncertainty on PDFs is estimated following the procedure of Ref.\ \cite{cteq6m} by reweighting the MC simulation according to each of the 20 pairs of error eigenvectors of the CTEQ6M PDF, with their effects added in quadrature.

\subsection{Modeling of detector} 
Uncertainties on the modeling of the detector include uncertainties on trigger efficiency, lepton identification and $b$-quark identification. The uncertainty on trigger efficiency is roughly 2.5\% for harder collisions [\ptt$ >90$ GeV or \mTT$ > 500$ GeV] and 6\% for softer collisions that are typically closer to trigger thresholds. The \ptt and \mTT differential cross sections are modified according to these uncertainties, and the \aetat differential cross section is rederived with trigger efficiencies reweighted according to \ptt. The identification efficiencies for $b$, $c$, light quarks ($u,d,s$) and gluons in MC simulations are calibrated using dijet data \cite{topmass_sdc}, and variations within the calibration uncertainty are used to determine the systematic uncertainty due to $b$-quark identification. Additional uncertainties arise from track multiplicity requirements on the selected jets in the identification of $b$ quarks. \\

Other instrumental uncertainties from modeling the detector arise from the calibration of the jet energy, resolution and efficiency. The jet energy scale (JES) corrects the measured energy of the jet to the energy of its constituent particles. The JES is derived using a quark-jet dominated $\gamma$ + jet sample, and corrects for the difference in detector response between data and simulation. An additional correction based on the single particle response accounts for the different characteristics of quark and gluon jets. Jets in MC simulations have their transverse momenta smeared so that the simulated resolution matches the one observed in data. Calibrations to the jet reconstruction and identification efficiency in MC simulations are determined using \zplus data. As mentioned earlier, jets are required to contain at least two tracks (see Sec.\ \ref{toc:data_samples}), and in MC simulations the corresponding efficiency is adjusted to match the one derived in dijet data. The uncertainties on the calibration of the jet energies, resolutions, and efficiencies as well as on the single particle response corrections are propagated to determine their effect on the differential cross sections.

\subsection{Sample composition} 
Uncertainties on the composition of the selected events arise from the
heavy-flavor scale factor used for \wplus events, the assumed \ttbar cross section, single top quark and diboson cross sections, and the estimate of the contributions from misidentified leptons. As described in Sec.\ \ref{toc:sampleCompSec}, the heavy-flavor scale factor in \wplus and the assumed \ttbar cross section are obtained from a simultaneous fit to the MVD distribution in the \lplustw, \lplusth and \lplusgefo samples. From the fit we derive a systematic uncertainty of 8\% on the normalization of the $Wc\bar{c}+\mm{jets}$ and $Wb\bar{b}+\mm{jets}$ processes, and 5\% on the normalization of the \ttbar processes. The uncertainty on the single top quark cross sections is 12.6\%, taken from varying the scale by factors of $2$ and $0.5$. An uncertainty of 7\% on the diboson cross sections is assigned to
the NLO predictions based on scale variation and PDF uncertainties. The  uncertainties on the data-driven method of estimating multijet (MJ) background and its kinematic dependencies, mostly due to the uncertainties on the selection rates of true and false lepton candidates, are 75\% in the \muplus and 32\% in the \eplus sample. These uncertainties are estimated by varying the contribution of $Wc\bar{c}+\mm{jets}$, $Wb\bar{b}+\mm{jets}$, $Zc\bar{c}+\mm{jets}$ and $Zb\bar{b}+\mm{jets}$ by $\pm 20\%$, the \ttbar contribution by $\pm 10\%$, comparing the fake and true signal rates in different variables (quoting the largest difference as additional parametrization uncertainty). In addition, to estimate the
contribution of the fake rate uncertainty, a different \met cut of $< 15$ GeV (standard cut for the fake rate estimation is $< 10$ GeV) \cite{DanielsMThesis} is applied. An overall 6.1\% uncertainty on the luminosity \cite{lumi_nim} is assigned to the measured cross sections and is fully correlated across all bins of the differential cross section.

\subsection{Regularization strength}
\label{toc:xsec_sys_reg} 
As a procedural uncertainty in the unfolding method, the regularization strength is changed to higher and lower values by amounts consistent with the general bounds discussed in Sec.\ \ref{toc:unfolding_intro}, and its impact is added to the total uncertainty. We test for a potential bias by doing a closure test employing an ensemble of simulated pseudo--data sets, and find biases smaller than the assigned systematic uncertainty due to the unfolding procedure.\\

%%%%%%%%%%%%%%%%%%%%%%%%%%%%%%%%%%%%%%%%%%%%%%%%%%%%%%%%%%%%%%%%%%%%%%%%
%%%%%%%%%%%%%%%%%%%%%%%%%%%%%%%%%%%%%%%%%%%%%%%%%%%%%%%%%%%%%%%%%%%%%%%%
%%%%%%%%%%%%%%%%%%%%%%%%%%%%%%%%%%%%%%%%%%%%%%%%%%%%%%%%%%%%%%%%%%%%%%%%

\section{Cross Sections}
\label{toc:results}
The inclusive \ttbar production cross section in the \ljets decay channel can be calculated from any of the three differential measurements. We calculate it from the average of the three differential measurements in events with $\ge4$ jets weighted by the $\chi^2$ as provided by the regularized unfolding (see Sec.\ \ref{toc:unfolding_intro}), and we find
\begin{equation}
\sigma^{t\bar{t}} = 8.0 \pm 0.7\thinspace(\mm{stat.}) \pm 0.6\thinspace(\mm{syst.}) \pm 0.5\thinspace(\mm{lumi.})~\mm{pb}.
\label{eqn:xsecResult}
\end{equation}
The inclusive \ttbar production cross sections using the individual differential cross sections in $d\sigma /dp_T^{\mm{top}}$, \aetat and \mTT are $8.0 \pm 1.1\thinspace(\mm{tot.})$ pb, $8.2 \pm 1.1\thinspace(\mm{tot.})$ pb and $7.8 \pm 1.0\thinspace(\mm{tot.})$ pb, respectively. The differences between these results have been verified to be statistically consistent using ensemble tests including correlations between the three measurements. These results are in agreement with the inclusive result of Sec.\ \ref{toc:sampleCompSec}, which was based on the inclusive \lplustw sample. The inclusive \ttbar production cross section [Eq.\ (\ref{eqn:xsecResult})] is in agreement with the inclusive fully resummed NNLL at NNLO QCD calculation (see Sec.\ \ref{toc:generators}), which gives $\sigma_{\mathrm{tot}}^{\mathrm{res}} = 7.35 ^{+0.23}_{-0.27}\thinspace(\mathrm{scale} + \mathrm{pdf})$ pb. The total cross section of the approximate NNLO calculation as in Refs.\ \cite{ptt_nnlo, etat_nnlo} is calculated from the \ptt distribution and yields $7.08 ^{+0.20}_{-0.24}\thinspace(\mathrm{scale}) ^{+0.36}_{-0.27} (\mathrm{PDF})$ pb. The data may also be compared to differential cross section predictions from \mcatnlo and \alpgen that correspond to total cross sections of $\sigma_{\mathrm{tot}} = 7.54$ pb and $\sigma_{\mathrm{tot}} = 5.61$ pb, respectively.

\begin{figure*}[ht]
\begin{center}
  \includegraphics[width=1.975\columnwidth,angle=0]{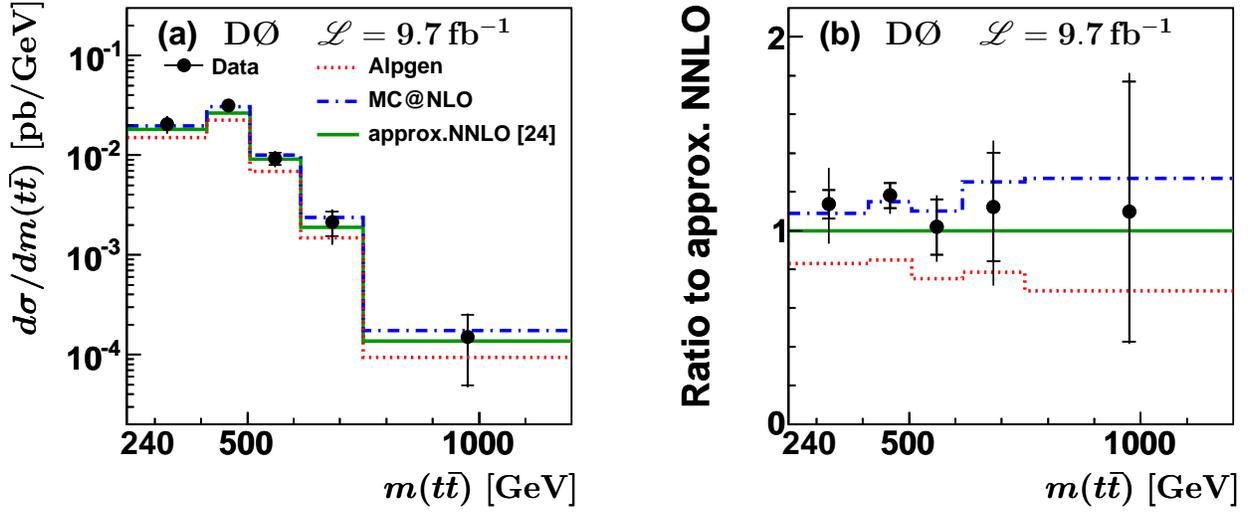}
\end{center}
\caption{(a) Measured differential cross section as a function of \mTT for data compared to several QCD predictions. The inner error bars correspond to the statistical uncertainties and the outer error bars to the total uncertainties. (b) Ratio of data, \alpgen (dashed line) and \mcatnlo cross sections (dash-dotted line) to the QCD prediction at approximate NNLO \cite{mtt_nnlo}. MC simulations and pQCD predictions use a top quark mass of $172.5$ GeV unless indicated to the contrary. Note that the correlated overall normalization uncertainty on the differential data points is about $\pm 6.6$\%.}
\label{fig:mTTxsec}
\end{figure*}

\begin{figure*}[htb]
\begin{center}
  \includegraphics[width=1.975\columnwidth,angle=0]{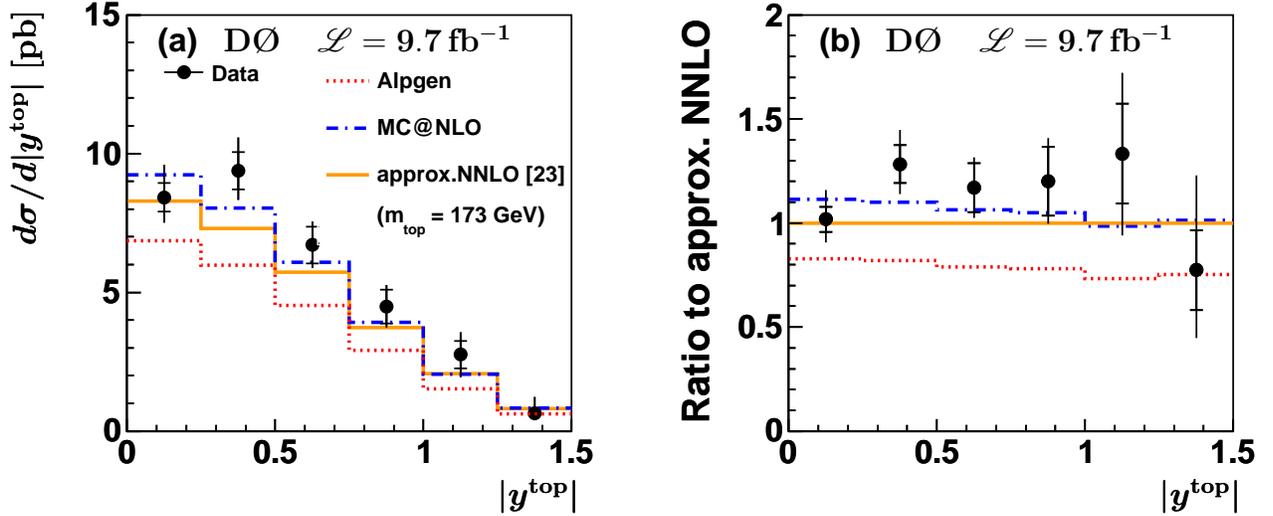}
\end{center}
 \caption{(a) Measured differential cross section as a function of \aetat for data compared to several QCD predictions. The inner error bars correspond to the statistical uncertainties and the outer error bars to the total uncertainties. (b) Ratio of data, \alpgen (dashed line) and \mcatnlo cross sections (dash-dotted line) to the QCD prediction at approximate NNLO \cite{etat_nnlo}. MC simulations and pQCD predictions use a top quark mass of $172.5$ GeV unless indicated to the contrary. Note that the correlated overall normalization uncertainty on the differential data points is about $\pm 6.6$\%.}
\label{fig:topaetaxsec}
\end{figure*}

\begin{figure*}[htb]
\begin{center}
  \includegraphics[width=1.975\columnwidth,angle=0]{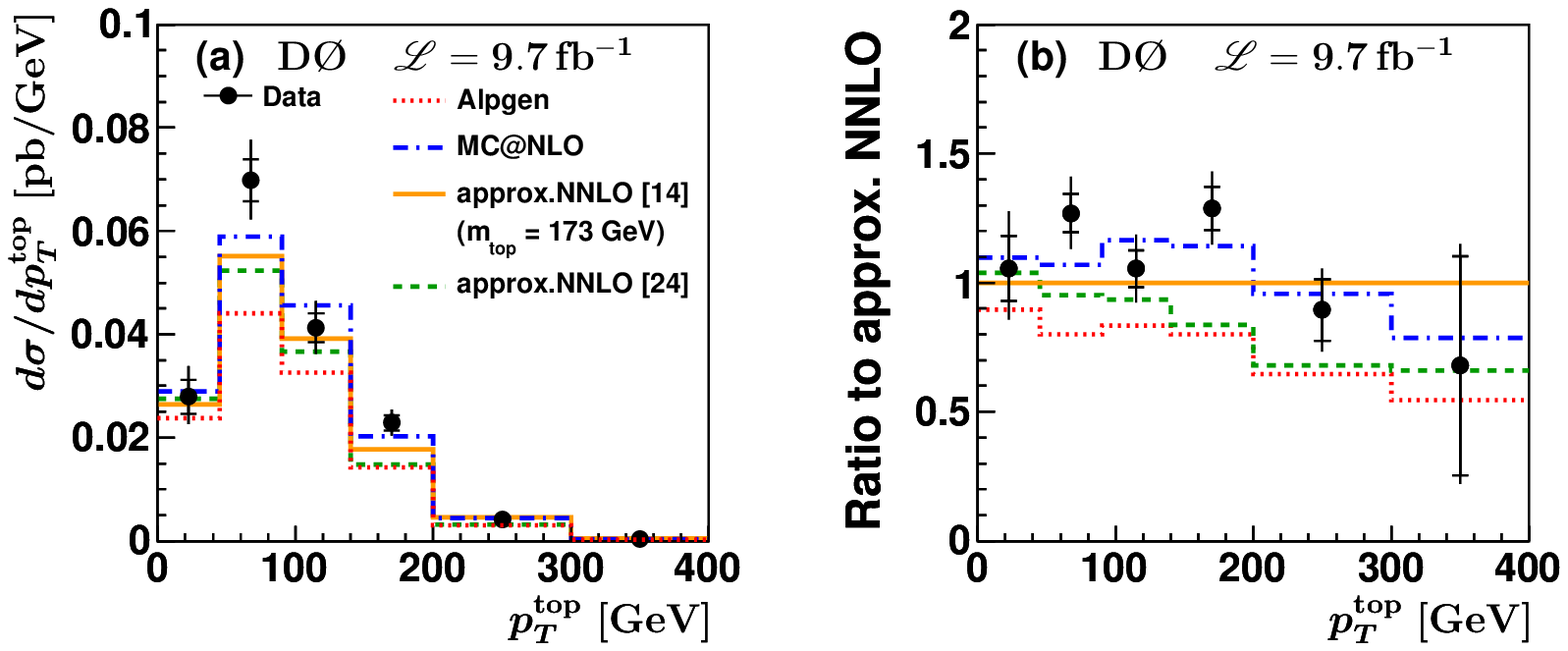}
\end{center}
\caption{(a) Measured differential cross section as a function of \ptt for data compared to several QCD predictions. The inner error bars correspond to the statistical uncertainties and the outer error bars to the total uncertainties. (b) Ratio of data, \alpgen (dashed line) and \mcatnlo cross sections (dash-dotted line) to the QCD prediction at approximate NNLO \cite{ptt_nnlo}. MC simulations and pQCD predictions use a top quark mass of $172.5$ GeV unless indicated to the contrary. Note that the correlated overall normalization uncertainty on the differential data points is about $\pm 6.6$\%.}
\label{fig:topptxsec}
\end{figure*}

The fully corrected differential cross sections are shown in Figs.\ \ref{fig:mTTxsec}--\ref{fig:topptxsec}, for \mTT, \aetat, and \ptt, respectively. The corresponding correlation coefficients of the differential measurements are presented in Tables \ref{tab:corr_mtt} to \ref{tab:corr_ptt} in Appendix \ref{toc:matrices}. For \ptt and \aetat distributions we present the average $t$ and $\bar{t}$ cross sections. The differential cross sections are listed in Table \ref{tab:xsec_mtt} to \ref{tab:xsec_ptt} in Appendix \ref{toc:matrices}. Note that the correlated normalization uncertainty on the differential data points is about $\pm 6.6$\%, dominated by the uncertainty on the measurement of the integrated luminosity. For quantitative comparison to SM predictions, the covariance matrices (Tables \ref{tab:cov_mtt}--\ref{tab:cov_ptt}) for the results are presented in Appendix \ref{toc:matrices}. No bin centering correction is applied to the measurements, and the cross sections are displayed at the center of each bin. Contributions beyond the highest bin boundary are included in the last bin of the \mTT, \aetat, and \ptt distributions. As shown in Fig. \ref{fig:mTTcontrolBg}, there are no contributions to the differential cross section for \mTT below $240$ GeV. 

Figure \ref{fig:mTTxsec}(a) shows the cross section for the unfolded data as a function of \mTT, and (b) shows the ratio of the cross section and several predictions to the approximate NNLO distribution \cite{mtt_nnlo}.
Within the systematic uncertainties the \mcatnlo and approximate NNLO describe the data, while the \alpgen prediction is low in absolute normalization as shown in Fig.\ \ref{fig:mTTxsec}(b). The distribution for \aetat is shown in Fig.\ \ref{fig:topaetaxsec}. The ratio in Fig.\ \ref{fig:topaetaxsec}(b) indicates that the distribution predicted by QCD at approximate NNLO is in marginal agreement with the data for \aetat. The predictions by \mcatnlo describe the data better.
As shown in Fig.\ \ref{fig:topptxsec}(a), the differential cross section as a function of \ptt is reasonably described by \mcatnlo and the approximate NNLO QCD prediction. The \mcatnlo prediction describes the shape of the \ptt distribution well.

This new result is consistent with an earlier measurement by \dzero using $1.0~\mm{fb^{-1}}$ of data \cite{dzero_ptt}. Statistical uncertainties are defined differently in Ref.\,\cite{dzero_ptt}, following Ref.\,\cite{GURU}, and are not directly comparable with the current uncertainties. The statistical uncertainties reported here are computed analytically and verified using an ensemble of simulated pseudo--data sets. Results presented here supersede the results of Ref.\ \cite{dzero_ptt}.

%%%%%%%%%%%%%%%%%%%%%%%%%%%%%%%%%%%%%%%%%%%%%%%%%%%%%%%%%%%%%%%%%%%%%%%%
%%%%%%%%%%%%%%%%%%%%%%%%%%%%%%%%%%%%%%%%%%%%%%%%%%%%%%%%%%%%%%%%%%%%%%%%
%%%%%%%%%%%%%%%%%%%%%%%%%%%%%%%%%%%%%%%%%%%%%%%%%%%%%%%%%%%%%%%%%%%%%%%%
\section{Conclusions}
Differential cross sections for \ttbar production have been measured in the \ljets decay channels using the full Tevatron data set at $\sqrt{s} = 1.96$ TeV. The data are corrected for detector efficiency, acceptance and bin migration by means of a regularized unfolding procedure. The differential cross sections are measured with a typical precision of $9\%$ as a function of the invariant mass of the \ttbar system \mTT, the absolute rapidity of the $t$ and $\bar{t}$ quarks \aetat, and the transverse momentum \ptt. The measured differential cross sections are in general agreement with predictions by QCD generators and predictions at approximate NNLO.

\include{New_acknowledgement_APS_full_names.tex}
% \section{Acknowledgments}
% % acknowledgement.tex                            28 August 2013
% %
We would like to thank W.\ Bernreuther and Z.\ G.\ Si for useful discussions on differential top quark cross sections. We thank the staffs at Fermilab and collaborating institutions, and acknowledge support from the
Department of Energy and National Science Foundation (United States of America);
Alternative Energies and Atomic Energy Commission and
National Center for Scientific Research/National Institute of Nuclear and Particle Physics  (France);
Ministry of Education and Science of the Russian Federation, 
National Research Center ``Kurchatov Institute" of the Russian Federation, and 
Russian Foundation for Basic Research  (Russia);
National Council for the Development of Science and Technology and
Carlos Chagas Filho Foundation for the Support of Research in the State of Rio de Janeiro (Brazil);
Department of Atomic Energy and Department of Science and Technology (India);
Administrative Department of Science, Technology and Innovation (Colombia);
National Council of Science and Technology (Mexico);
National Research Foundation of Korea (Korea);
Foundation for Fundamental Research on Matter (The Netherlands);
Science and Technology Facilities Council and The Royal Society (United Kingdom);
Ministry of Education, Youth and Sports (Czech Republic);
Bundesministerium f\"{u}r Bildung und Forschung (Federal Ministry of Education and Research) and 
Deutsche Forschungsgemeinschaft (German Research Foundation) (Germany);
Science Foundation Ireland (Ireland);
Swedish Research Council (Sweden);
China Academy of Sciences and National Natural Science Foundation of China (China);
and
Ministry of Education and Science of Ukraine (Ukraine).

% We thank the staffs at Fermilab and collaborating institutions,
% and acknowledge support from the
% DOE and NSF (USA);
% CEA and CNRS/IN2P3 (France);
% MON, NRC KI and RFBR (Russia);
% CNPq, FAPERJ, FAPESP and FUNDUNESP (Brazil);
% DAE and DST (India);
% Colciencias (Colombia);
% CONACyT (Mexico);
% NRF (Korea);
% FOM (The Netherlands);
% STFC and the Royal Society (United Kingdom);
% MSMT and GACR (Czech Republic);
% BMBF and DFG (Germany);
% SFI (Ireland);
% The Swedish Research Council (Sweden);
% and
% CAS and CNSF (China).
%

% \newpage{}

%%%%%%%%%%%%%%%%%%%%%%%%%%%%%%%%%%%%%%%%%%%%%%%%%%%%%%%%%%%%%%%%%%%%%%%%
%%%%%%%%%%%%%%%%%%%%%%%%%%%%%%%%%%%%%%%%%%%%%%%%%%%%%%%%%%%%%%%%%%%%%%%%
%%%%%%%%%%%%%%%%%%%%%%%%%%%%%%%%%%%%%%%%%%%%%%%%%%%%%%%%%%%%%%%%%%%%%%%%
\appendix

\section[Results of the Unfolding]{Cross section tables and covariance matrices}
\label{toc:matrices}

The correlation coefficients for the differential cross sections are given in Tables \ref{tab:corr_mtt}, \ref{tab:corr_aetat}, and \ref{tab:corr_ptt}, which are helpful in interpreting the differential cross sections as shown in Figs.\ \ref{fig:mTTxsec}, \ref{fig:topaetaxsec}, and \ref{fig:topptxsec}. The numerical values of the cross sections are given as a function of \mTT, \aetat, and \ptt in Table \ref{tab:xsec_mtt}, \ref{tab:xsec_aetat}, and \ref{tab:xsec_ptt}, respectively. Contributions beyond the highest bin boundary are included in the last bin of the \mTT, \aetat, and \ptt table entries. The full covariance matrices for these cross sections are given in Tables \ref{tab:cov_mtt}, \ref{tab:cov_aetat}, and \ref{tab:cov_ptt}. The results of diagonalizing the covariance matrices in terms of eigenvalues and corresponding eigenvectors are presented in Tables \ref{tab:eigen_mtt}, \ref{tab:eigen_aetat}, and \ref{tab:eigen_ptt}.

\begin {table*}[th]%
\centering %
\caption {\label{tab:corr_mtt} Correlation coefficients of the differential cross section as a function of \mTT.}
\begin{ruledtabular}
\begin {tabular}{lc@{\hskip 2.4ex}c@{\hskip 2.4ex}c@{\hskip 2.4ex}c@{\hskip 2.4ex}c}
 \mTT [TeV] &  $0.2400$ -- $0.4125$ &  $0.4125$ -- $0.5050$ & $0.5050$ -- $0.6150$ & $0.6150$ -- $0.7500$ & $0.7500$ -- $1.200$ \T \\ \hline
 $0.2400$ -- $0.4125$ & $1  $     &$-0.45$ & $+0.13$ &$-0.02$ &$-0.00 $\T \\ 
 $0.4125$ -- $0.5050$ &$-0.45$ & $1 $      &$-0.51$ &$+0.12$ &$+0.01 $\\
 $0.5050$ -- $0.6150$ &$+0.13$ &$-0.51$ & $1 $      &$-0.48$ &$+0.02$ \\
 $0.6150$ -- $0.7500$ &$-0.02 $& $+0.12$ &$-0.48$ & $1 $      &$-0.63$ \\
 $0.7500$ -- $1.2000$  &$-0.00 $& $+0.01$ &$+0.02 $&$-0.63$ & $1  $     \\
\end {tabular}
\end{ruledtabular}
\end {table*}

\begin {table*}[th]%
\centering %
\caption {\label{tab:corr_aetat} Correlation coefficients of the differential cross section as a function of \aetat.}
\begin{ruledtabular}
\begin {tabular}{lc@{\hskip 2.4ex}c@{\hskip 2.4ex}c@{\hskip 2.4ex}c@{\hskip 2.4ex}c@{\hskip 2.4ex}c}
 \aetat &  $0.00$ -- $0.25$ &  $0.25$ -- $0.50$ & $0.50$ -- $0.75$ & $0.75$ -- $1.00$ & $1.00$ -- $1.25$ & $1.25$ -- $1.50$ \T \\ \hline
 $0.00$ -- $0.25$   & $1 $   &$-0.51$ &$-0.06 $&$-0.02$ &$-0.01$ &$-0.00 $\T \\ 
 $0.25$ -- $0.50$   &$-0.51$ & $1  $  &$-0.39$ &$-0.02$ &$-0.01$ &$-0.01 $\\
 $0.50$ -- $0.75$   &$-0.06$ &$-0.39$ & $1   $ &$-0.41 $&$-0.00$ &$-0.00 $\\
 $0.75$ -- $1.00$   &$-0.02$ &$-0.02$ &$-0.41 $& $1 $  & $-0.41$ &$-0.01 $\\
 $1.00$ -- $1.25$   &$-0.01$ &$-0.01$ &$-0.00 $&$-0.41 $&$1 $    &$-0.46 $\\
 $1.25$ -- $1.50$   &$-0.00$ &$-0.01$ &$-0.00 $&$-0.01 $&$-0.46$ & $1     $  \\
\end {tabular}
\end{ruledtabular}
\end {table*}

\begin {table*}[th]%
\centering %
\caption {\label{tab:corr_ptt} Correlation coefficients of the differential cross section as a function of \ptt.}
\begin{ruledtabular}
\begin {tabular}{lc@{\hskip 2.4ex}c@{\hskip 2.4ex}c@{\hskip 2.4ex}c@{\hskip 2.4ex}c@{\hskip 2.4ex}c}
 \ptt [TeV] &  $0.000$ -- $0.045$ &  $0.045$ -- $0.090$ & $0.090$ -- $0.140$ & $0.140$ -- $0.200$ & $0.200$ -- $0.300$ & $0.300$ -- $0.500$ \T \\ \hline
 $0.000$ -- $0.045$   &$1    $   &$-0.55 $&$+0.01$ &$+0.00 $&$-0.00$ &$-0.00 $\T \\ 
 $0.045$ -- $0.090$   &$-0.55 $& $1   $    &$-0.42$ &$+0.02 $&$+0.00$ &$-0.00 $\\
 $0.090$ -- $0.140$   &$+0.01 $&$-0.42 $&$1   $    &$-0.37 $&$-0.01$ &$-0.00 $\\
 $0.140$ -- $0.200$   &$+0.00 $&$+0.02 $&$-0.37 $& $1   $    &$-0.29 $&$-0.03$ \\
 $0.200$ -- $0.300$   &$-0.00 $&$+0.00$ &$-0.01$ &$-0.29 $& $1    $   &$-0.15 $\\
 $0.300$ -- $0.500$   &$-0.00$ &$+0.00 $&$-0.00 $&$-0.03 $&$-0.15 $& $1 $      \\
\end {tabular}
\end{ruledtabular}
\end {table*}

\begin {table*}[t]%
\centering %
\caption {\label{tab:xsec_mtt} Average value of \mTT and differential cross section in each bin of \mTT. In addition to the systematic uncertainty reported in column five there is a 6.1\% normalization uncertainty across all bins due to the uncertainty on the integrated luminosity.}
\begin{ruledtabular}
\begin {tabular}{lcccc}
\mTT [TeV] & $\langle M(t\bar{t}) \rangle$ [TeV]  & $d\sigma/dM(t\bar{t}) \mm{[pb/TeV]}$ & $\delta^{\mm{stat.}} \mm{[pb/TeV]}$ & $\delta^{\mm{sys.}} \mm{[pb/TeV]}$ \T \\ \hline
 $0.2400$ -- $0.4125$       & $0.36$ & $20.60$    & $\pm 1.52$        & $^{+3.86}_{-3.76}$ \T \\
 $0.4125$ -- $0.5050$       & $0.46$ & $31.26$    & $\pm 2.03$        & $^{+1.84}_{-2.20}$ \T \\
 $0.5050$ -- $0.6150$       & $0.55$ & $9.38$    & $\pm 1.34$        & $^{+0.78}_{-1.00}$ \T \\
 $0.6150$ -- $0.7500$       & $0.67$ & $2.13$    & $\pm 0.59$        & $^{+0.43}_{-0.63}$ \T \\
 $0.7500$ -- $1.2000$        & $0.83$ & $0.15$    & $\pm 0.10$        & $^{+0.06}_{-0.05}$ \T \\
\end {tabular}
\end{ruledtabular}
\end {table*}

\begin {table*}[t]%
\centering %
\caption {\label{tab:xsec_aetat} Average value of \aetat and differential cross section in each bin of \aetat. In addition to the systematic uncertainty reported in column five there is a 6.1\% normalization uncertainty across all bins due to the uncertainty on the integrated luminosity.}
\begin{ruledtabular}
\begin {tabular}{lcccc}
\aetat & $\langle |y|(t/\bar{t}) \rangle$  & $d\sigma/d|y|(t/\bar{t}) \mm{[pb]}$ & $\delta^{\mm{stat.}} \mm{[pb]}$ & $\delta^{\mm{sys.}} \mm{[pb]}$ \T \\ \hline
 $0.00$ -- $0.25$   &$0.13$ & $8.50$    & $\pm 0.51$        & $^{+0.67}_{-0.99}$ \T \\
 $0.25$ -- $0.50$   &$0.37$ & $9.46$    & $\pm 0.67$        & $^{+0.63}_{-0.88}$ \T \\
 $0.50$ -- $0.75$   &$0.62$ & $6.72$    & $\pm 0.67$        & $^{+0.29}_{-0.30}$ \T \\
 $0.75$ -- $1.00$   &$0.86$ & $4.64$    & $\pm 0.64$        & $^{+0.36}_{-0.41}$ \T \\
 $1.00$ -- $1.25$   &$1.11$ & $2.73$    & $\pm 0.49$        & $^{+0.66}_{-0.63}$ \T \\
 $1.25$ -- $1.50$   &$1.36$ & $0.63$    & $\pm 0.16$        & $^{+0.25}_{-0.25}$ \T \\
\end {tabular}
\end{ruledtabular}
\end {table*}

\begin {table*}[t]%
\centering %
\caption {\label{tab:xsec_ptt} Average value of \ptt and differential cross section in each bin of \ptt. In addition to the systematic uncertainty reported in column five there is a 6.1\% normalization uncertainty across all bins due to the uncertainty on the integrated luminosity.}
\begin{ruledtabular}
\begin {tabular}{lcccc}
\ptt [TeV] & $\langle p_{T}(t/\bar{t}) \rangle$ [TeV]  &  $d\sigma/dp_{T}(t/\bar{t}) \mm{[pb/TeV]}$ & $\delta^{\mm{stat.}} \mm{[pb/TeV]}$ & $\delta^{\mm{sys.}} \mm{[pb/TeV]}$ \T \\ \hline
 $0.000$ -- $0.045$       &$0.030$ & $27.76$    & $\pm 3.31$        & $^{+3.21}_{-4.29}$ \T \\
 $0.045$ -- $0.090$       &$0.068$ & $69.70$    & $\pm 4.07$        & $^{+1.79}_{-2.88}$ \T \\
 $0.090$ -- $0.140$       &$0.112$ & $41.47$    & $\pm 2.78$        & $^{+3.34}_{-3.45}$ \T \\
 $0.140$ -- $0.200$       &$0.164$ & $22.84$    & $\pm 1.51$        & $^{+1.25}_{-1.34}$ \T \\
 $0.200$ -- $0.300$       &$0.234$ & $ 4.18$    & $\pm 0.56$        & $^{+0.41}_{-0.39}$ \T \\
 $0.300$ -- $0.500$       &$0.321$ & $ 0.32$    & $\pm 0.20$        & $^{+0.07}_{-0.09}$ \T \\
\end {tabular}
\end{ruledtabular}
\end {table*}

\begin {table*}[th]%
\centering %
\caption {\label{tab:cov_mtt} Covariance matrix (statistical and systematical uncertainties) of the differential cross section as a function of \mTT. The systematic uncertainty is assumed to be 100\% correlated.}
\begin{ruledtabular}
\begin {tabular}{lc@{\hskip 2.4ex}c@{\hskip 2.4ex}c@{\hskip 2.4ex}c@{\hskip 2.4ex}c}
 \mTT [TeV] &  $0.2400$ -- $0.4125$ &  $0.4125$ -- $0.5050$ & $0.5050$ -- $0.6150$ & $0.6150$ -- $0.7500$ & $0.7500$ -- $1.200$\\ \hline
 $0.2400$ -- $0.4125$ &$+16.832$ &$-1.430$  &$+ 0.364$ &$-0.051$ &$-0.001$ \T \\ 
 $0.4125$ -- $0.5050$ &$-1.430$   & $+6.436$  &$-1.820$ &$+ 0.321 $&$+ 0.021$   \\
 $0.5050$ -- $0.6150$ &$+0.364$   &$-1.820$  & $+2.570$ &$-0.635$ &$+ 0.020 $  \\
 $0.6150$ -- $0.7500$ &$-0.051$   &$+0.321$  &$-0.635$ & $+0.633$ &$-0.141$  \\
 $0.7500$ -- $1.2000$  &$-0.001$   &$+0.021 $ & $+0.020$ &$-0.141$ & $+0.129 $ \\
\end {tabular}
\end{ruledtabular}
\end {table*}

\begin {table*}[th]%
\centering %
\caption {\label{tab:cov_aetat} Covariance matrix (statistical and systematical uncertainties) of the differential cross section as a function of \aetat. The systematic uncertainty is assumed to be 100\% correlated.}
\begin{ruledtabular}
\begin {tabular}{lc@{\hskip 2.4ex}c@{\hskip 2.4ex}c@{\hskip 2.4ex}c@{\hskip 2.4ex}c@{\hskip 2.4ex}c}
 \aetat &  $0.00$ -- $0.25$ &  $0.25$ -- $0.50$ & $0.50$ -- $0.75$ & $0.75$ -- $1.00$ & $1.00$ -- $1.25$ & $1.25$ -- $1.50$\\ \hline
 $0.00$ -- $0.25$   &$+0.952$ &$-0.164$ &$-0.017$ &$-0.004$ &$-0.001$ &$-0.000$ \T \\ 
 $0.25$ -- $0.50$   &$-0.164$ &$+1.029$ &$-0.163$ &$-0.008$ &$-0.001$ &$-0.001$ \\
 $0.50$ -- $0.75$   &$-0.017$ &$-0.163$ &$+0.551$ &$-0.155$ &$-0.001$ &$-0.000$ \\
 $0.75$ -- $1.00$   &$-0.004$ &$-0.008$ &$-0.155$ &$+0.557$ &$-0.121$ &$-0.002$ \\
 $1.00$ -- $1.25$   &$-0.001$ &$-0.001$ &$-0.001$ &$-0.121$ &$+0.609$ &$-0.062$ \\
 $1.25$ -- $1.50$   &$-0.000$ &$-0.001$ &$-0.000$ &$-0.002$ &$-0.062$ &$+0.087$ \\
\end {tabular}
\end{ruledtabular}
\end {table*}

\begin {table*}[th]%
\centering %
\caption {\label{tab:cov_ptt} Covariance matrix (statistical and systematical uncertainties) of the differential cross section as a function of \ptt. The systematic uncertainty is assumed to be 100\% correlated.}
\begin{ruledtabular}
\begin {tabular}{lc@{\hskip 2.4ex}c@{\hskip 2.4ex}c@{\hskip 2.4ex}c@{\hskip 2.4ex}c@{\hskip 2.4ex}c}
 \ptt [TeV] &  $0.000$ -- $0.045$ &  $0.045$ -- $0.090$ & $0.090$ -- $0.140$ & $0.140$ -- $0.200$ & $0.200$ -- $0.300$ & $0.300$ -- $0.500$ \\ \hline
 $0.000$ -- $0.045$   &$+25.018$ &$-8.692 $ &$+0.157 $ & $+0.011 $&$-0.008 $&$-0.000 $\T \\ 
 $0.045$ -- $0.090$   &$-8.692$  &$+22.028$ &$-5.916$  & $+0.155 $&$+0.0149 $&$+0.000$ \\
 $0.090$ -- $0.140$   &$+0.157$  &$-5.916$  &$+19.277$ &$-1.958 $&$-0.037 $&$-0.001$ \\
 $0.140$ -- $0.200$   &$+0.011 $ &$+0.155 $ &$-1.958 $ &$+3.942 $&$-0.324 $&$-0.009 $\\
 $0.200$ -- $0.300$   &$-0.008$  &$+0.015 $ &$-0.037$  &$-0.324 $&$+0.469 $&$-0.013 $\\
 $0.300$ -- $0.500$   &$-0.000$  &$+0.000 $ &$-0.001 $ &$-0.009 $&$-0.013 $&$+0.047$  \\
\end {tabular}
\end{ruledtabular}
\end {table*}

\begin {table*}[th]%
\centering %
\caption {\label{tab:eigen_mtt} Eigenvalues and eigenvectors of the covariance matrix (see Table \ref{tab:cov_mtt}) of the differential cross section as a function of \mTT. The contribution of the eigenvector is listed in the first column together with its error calculated as the square root of the eigenvalue $\lambda$. The eigenvalue $\lambda$ in the second column followed by the elements of the eigenvectors in bins of \mTT.}
\begin{ruledtabular}
\begin {tabular}{lcccccc}
\multicolumn{1}{l}{Contribution [pb/TeV]} & \multicolumn{1}{c}{$\lambda$} & \multicolumn{5}{c}{\mTT range [TeV]} \\
                  & & $0.2400$ -- $0.4125$ & $0.4125$ -- $0.5050$ & $0.5050$ -- $0.6150$ & $0.6150$ -- $0.7500$ & $0.7500$ -- $1.2000$ \\ \hline
$\hphantom{0}1.655 \pm 0.284$ & $\hphantom{0}0.081$ &$-0.000$ &$+0.003$ &$+0.079$ &$+0.330$ &$+0.941$ \\
$\hphantom{0}6.361 \pm 0.691$ & $\hphantom{0}0.478$ &$+0.000$ &$+0.050$ &$+0.316$ &$+0.886$ &$-0.337$ \\
$19.747 \pm 1.416$ & $\hphantom{0}2.004$ &$+0.015$ &$+0.383$ &$+0.867$ &$-0.316$ &$+0.037$ \\
$28.166 \pm 2.643$ & $\hphantom{0}6.985$ &$+0.147$ &$+0.911$ &$-0.375$ &$+0.082$ &$+0.000$\\
$16.360 \pm 4.129$ & $17.052$            &$+0.989$ &$-0.141$ &$+0.043$ &$-0.007$ &$-0.000$\\
\end {tabular}
\end{ruledtabular}
\end {table*}

\begin {table*}[th]%
\centering %
\caption {\label{tab:eigen_aetat} Eigenvalues and eigenvectors of the covariance matrix (see Table \ref{tab:cov_aetat}) of the differential cross section as a function of \aetat. The contribution of the eigenvector is listed in the first column together with its error calculated as the square root of the eigenvalue $\lambda$. The eigenvalue $\lambda$ in the second column followed by the elements of the eigenvectors in bins of \aetat.}
\begin{ruledtabular}
\begin {tabular}{lccccccc}
\multicolumn{1}{l}{Contribution [pb]} & \multicolumn{1}{c}{$\lambda$} & \multicolumn{6}{c}{\aetat range} \\
                  & & $0.00$ -- $0.25$ &$0.25$ -- $0.50$ &$0.50$ -- $0.75$ &$0.75$ -- $1.00$ &$1.00$ -- $1.25$ &$1.25$ -- $1.50$ \\ \hline
$\hphantom{0}0.387 \pm 0.283$ & 0.080 &$-0.000 $& $ +0.001 $& $-0.010 $&$+0.027 $& $-0.111 $& $+0.993$ \\
$\hphantom{0}0.934 \pm 0.590$ & 0.348 &$+0.029 $&$-0.164 $&$+0.685 $&$-0.662 $&$+0.248 $&$+0.053 $\\
$\hphantom{0}2.496 \pm 0.763$ & 0.582 &$+0.080$ &$ -0.249 $&$+0.587 $&$ +0.430$ &$ -0.630 $&$ -0.076 $\\
$\hphantom{0}5.194 \pm 0.872$ & 0.761 &$+0.200 $ &$-0.273 $&$+0.214 $ &$+0.570 $ &$ +0.715 $&$ +0.067 $\\
$\hphantom{0}0.864 \pm 0.922$ & 0.851 &$-0.800 $&$+0.455 $&$+0.297$ &$+0.212 $&$+0.138 $&$+0.012$ \\
$           14.188 \pm 1.092$ & 1.192 &$+0.559$&$+0.794$&$+0.229 $&$+0.073$&$+0.020$ &$+0.002 $\\
\end {tabular}
\end{ruledtabular}
\end {table*}

\begin {table*}[th]%
\centering %
\caption {\label{tab:eigen_ptt} Eigenvalues and eigenvectors of the covariance matrix (see Table \ref{tab:cov_ptt}) of the differential cross section as a function of \ptt. The contribution of the eigenvector is listed in the first column together with its error calculated as the square root of the eigenvalue $\lambda$. The eigenvalue $\lambda$ in the second column followed by the elements of the eigenvectors in bins of \ptt.}
\begin{ruledtabular}
\begin {tabular}{lccccccc}
\multicolumn{1}{l}{Contribution [pb/TeV]} & \multicolumn{1}{c}{$\lambda$} & \multicolumn{6}{c}{\ptt range [TeV]} \\
                  & & $0.000$ -- $0.045$ &$0.045$ -- $0.090$ &$0.090$ -- $0.140$& $0.140$ -- $0.200$ &$0.200$ -- $0.300$ &$0.300$ -- $0.500$ \\ \hline
$\hphantom{0}0.648 \pm 0.214$ & $\hphantom{0}0.046$ & $+0.000$&$+0.000$&$+0.001$&$+0.006$&$+0.036$&$+0.999$ \\
$\hphantom{0}7.162 \pm 0.661$ & $\hphantom{0}0.437$ &$ +0.001$&$ +0.003$&$+0.013$&$ +0.099$&$+0.994$&$-0.037$ \\
$31.477 \pm 1.924$ & $\hphantom{0}3.703$ &$ +0.017$&$ +0.045$&$+0.140$&$ +0.984$&$ -0.100$&$-0.002$ \\
$81.156 \pm 3.451$ & $11.906$&$+0.461$&$+0.705$&$+0.526$&$-0.115$&$+0.002$&$+0.000$\\ 
$\hphantom{0}1.400 \pm 4.601$ & $21.160$&$+0.561$ &$+0.235$&$-0.788$&$+0.092$&$-0.000$&$+0.000$\\
$16.028 \pm 5.790$ & $33.529$&$-0.687 $ &$+0.667 $& $-0.288 $ &$+0.022 $ &$+0.001 $& $+0.000 $\\
\end {tabular}
\end{ruledtabular}
\end {table*}

\end{document}